**Toward Valley-coupled Spin Qubits**

*Kuan Eng Johnson Goh[1,2,]\*, Fabio Bussolotti[1], Chit Siong Lau[1], Dharmraj Kotekar-Patil[1], Zi En Ooi[1], Jingyee Chee[1]*

[1] Institute of Materials Research and Engineering, Agency for Science Technology and Research, 2 Fusionopolis Way, 08-03 Innovis, 138634, Singapore

[2] Department of Physics, National University of Singapore, 2 Science Drive 3, 117551, Singapore

\* E-mail: kejgoh@yahoo.com; gohj@imre.a-star.edu.sg; phygkej@nus.edu.sg;

**Abstract**

The bid for scalable physical qubits has attracted many possible candidate platforms. In particular, spin-based qubits in solid-state form factors are attractive as they could potentially benefit from processes similar to those used for conventional semiconductor processing. However, material control is a significant challenge for solid-state spin qubits as residual spins from substrate, dielectric, electrodes or contaminants from processing contribute to spin decoherence. In the recent decade, valleytronics has seen a revival due to the discovery of valley-coupled spins in monolayer transition metal dichalcogenides. Such valley-coupled spins are protected by inversion asymmetry and time reversal symmetry and are promising candidates for robust qubits. In this report, the progress toward building such qubits is presented. Following an introduction to the key attractions in fabricating such qubits, an up-to-date brief is provided for the status of each key step, highlighting advancements made and/or outstanding work to be done. This report concludes with a perspective on future development highlighting major remaining milestones toward scalable spin-valley qubits.

# 1. Introduction

There is an intense race to scale up the number of qubits in the quantum computer for quantum advantage (or supremacy) ostensibly to demonstrate finally that the quantum computer is able to solve certain problems deemed infeasible for its classical counterpart.[1,2] While it is possible to expediently build the support stack for demonstrating one or two qubits using "off-the-shelf" components and standard measurement electronics, scaling up to a full-fledge universally programmable quantum computer would require the ability to not only scale the qubits, but also the concomitant scalable development of the peripheral components for quantum state readout, manipulation and transfer.[3–6] Increasing from a few tens of qubits to beyond thousands of qubits would require rather radical rethinking of the overall architecture, material and process compatibility, and operational strategy. The tour de force to build a quantum computer typically involve significant investments at the national level (e.g. Australia, Canada, US, EU, UK),[7–12] or by industry juggernauts,[13] since the entire technology stack needed to drive a quantum computer is quite specific to the particular type of qubit used. To this end, a solid-state qubit platform that could be compatible with the existing Si microelectronics is often seen as beneficial and economical since the classical electronics portion is already a well-established technology with ready foundries.

Silicon-based qubit technologies appear to be an obvious choice and various groups have pursued this option using the spin states associated with single dopant atoms[6,14–16] or electrostatically gated quantum dots[3,17,18] as qubits in Si. The key challenges in these strategies are the atomically precise dopant placement,[19] access and control of single dopant states,[20] and the negligible spin-orbit interaction in Si perhaps makes it less efficient for electrical access and control.[21]

More recently, 2D materials have emerged as an alternative class of materials which could be potential solid-state hosts for spin-based qubits.[22–25] In particular, single layer transition metal dichalcogenides (TMDCs) are 2D semiconductors with an intrinsic band-gap and strong spin-orbit coupling which promises fast spin operation times.[22,23,26] Significantly, the inherent inversion asymmetry in such TMDCs results in a unique spin-valley coupling which is expected to enhance the coherence lifetime of spin-valley states thus making the TMDC platform advantageous for robust qubits.[22,23,27] The single layer TMDCs are readily transferrable[28] to a Si wafer or may be grown[29] directly onto a Si wafer, and the majority of TMDC-based devices have been fabricated on the Si wafer.[30,31] Significant research efforts are also underway to develop strategies for compatibility with Si CMOS technologies.[32,33] Hence, the 2D TMDC family is a very compelling alternative for building solid-state qubits. Various groups[31,34–40] have therefore invested efforts (see for example the facilities highlighted in **Figure 1**) to build electrostatically gated quantum dots in 2D TMDCs which could potentially lead to the development of qubits on the same platform. This report provides an update of the progress in this field. The key challenges in materials development will be highlighted. Following a brief review of potential device architectures proposed in the literature, two quantum dot fabrication strategies are presented together with discussions on the respective experimental progress. This report concludes with a summary of the key achievements to date, the challenges ahead, and provides an outlook for the field.

## 2. Materials Development: Challenges and Progress

### 2.1. Growth of high quality TMDCs

TMDCs tend to occur naturally or grow preferentially in the 2H phase[41] which is semiconducting and non-centrosymmetric in monolayer form. The lack of an in-plane inversion center, implies that the single layer TMDCs intrinsically host inequivalent valleys at the K and K′ points of their hexagonal Brillouin Zone.[27] In addition, the large spin-orbit coupling carried by the transition atoms combined with the time reversal symmetry, introduces an energy splitting of opposite signs at the K and K′ valleys, leading to a coupling of opposite spins to the K and K′ valleys. This unique "spin-locking" mechanism makes the valleys addressable via their incumbent spins and primes the single layer TMDCs for spintronic and valleytronic applications, where the spin and valley degrees of freedom are explored for accelerating electronic computing and information processing.[27]

Various synthesis approaches and strategies were explored to obtain high quality large scale single- and few-layer TMDCs so to fully exploit their promising properties in various applications. TMDC layers with lateral size of few microns were first obtained from bulk single crystals by mechanical exfoliation. Despite the possibility of isolating high structural and electronic quality of small exfoliated TMDCs flakes, this technique is unlikely to be scalable for industrial production. To meet the wafer-scale fabrication requirements for industry adoption, different TMDC preparation methods are required. Among them, chemical vapor deposition (CVD) is a promising option, presently allowing to synthesize TMDCs triangular domains (10~100 μm) over a large (~mm size) substrate area. In its simplest form, the CVD growth process consists of the co-evaporation of metal oxides and chalcogen precursors that lead to vapor phase reaction followed by the formation of a stable TMDC layer over a suitable substrate.[42] Despite the considerable experimental and theoretical research efforts,[42] there remains major hurdles toward achieving high quality as grown TMDC layers. Chemical defects (i.e. vacancies) and grain boundaries introduced during the growth process, for example, can affect both the TMDC optical and electronic properties of the TMDC film.[43,44] As the nucleation density is a determinant of the film quality, various growth parameters can be adjusted to reduce the number of nucleation points, so as to achieve a continuous film made of a few larger single-crystal domains.[45] The atomic layer of TMDC materials is also influenced by the nanoscale surface morphology, terminating atomic planes of substrates, as well as by lattice mismatching between the layer and the substrate.[46] Recently, Lim et al.[30] succeeded in synthesizing highly oriented wafer-scale $MoS_2$ using a single crystal sapphire substrate. The synthesized $MoS_2$ showed a highly ordered in plane distribution with two main orientation (30º differences). This highly oriented $MoS_2$ film could also be easily removed from the sapphire substrate by simply immersing the substrate into water.[30] These results show that growth on a single crystal substrate could be another key factor toward the CVD growth of large-area, oriented high-quality TMDCs.

**2.2. TMDCs quality assessment by optical and electronic properties**

In current state-of-the-art valley physics experiments, specific valleys can be commonly targeted and selectively populated by controlling the helicity of the incident radiation on the material. This possibility of selective valley population was first demonstrated in monolayer $MoS_2$ through optical pumping of circularly polarized light[47–49] and then confirmed for a wider class of single layer TMDCs.[49–51] The opposite spin polarization at K and K′ also results in unlike spins moving in opposite directions, perpendicular to an in-plane electric field (valley Hall effect) which has been reported for single layer [52] and bilayer[53] $MoS_2$ based transistors.

Optical techniques such as microscopy and Raman mapping are frequently the first line of quality checks carried out for assessing TMDC flake size or growth coverage, and layer numbers.[54] For checking the TMDC specific band structure fingerprints, photoluminescence (PL), with linearly- or circularly-polarized light, is a fast-turnaround and non-invasive tool.[55] A major issue related to the determination of valley related lifetimes is the inconsistencies in material quality used in optical based experimental studies of TMDCs. In this context, a wide range of life-times (from picoseconds to nanoseconds) was reported depending of the excitation species (exciton, trions, etc.).[27] Analogous to other 2D materials, single layer TMDCs are in fact difficult to isolate from the environment, and thus determining their intrinsic properties can be quite challenging. The physical properties of 2D materials are often affected by unknown substrate interactions originating for example by the presence of interface defects.[56] Free-standing layers may also alter electronic and optical properties as a result of intrinsic structural relaxation such as rippling.[56] Finally, the temperature of the system can also directly control the valley polarization/depolarization in single layer TMDCs via phonon activated processes. This is well exemplified in the recent work reported by Chellappan et al. (reproduced in **Figure 2**) showing the circular dichroic PL response of a single layer WSe$_2$ at 8 K and 300 K.[57] In particular, this study observed a dramatic reduction of the valley polarization from 45% to 5% with increasing temperature. Therein, a detailed analysis of the emission line-width suggests that this change is caused by the strong exciton–phonon interactions which efficiently scatter the excitons into different excitonic states that are easily accessible due to the supply of excess photoexcitation energy.

As the unique spin-valley coupling is deemed a key advantage of using the TMDC for valleytronics and qubits development, the determination of this property via circular dichroic PL measurements would be an important first step in screening the TMDC materials before extensive device fabrication is carried out. Here, a dedicated circular dichroic PL system is useful to enable high-throughput screening for valley polarization in TMDC materials.[58] **Figure 3** demonstrates an example of such a screening done for a single-layer WS2 grown using CVD on a sapphire single-crystal substrate. The sample was mounted in a cryostat with an optical window and cooled to a minimum of ~4 K using a closed cycle helium circulation. Circularly polarized 594 nm laser light was focused onto the sample, and the resulting PL collected with the same microscope objective and split into two beams of orthogonal polarization states. The separated beams were then coupled into a spectrometer via a bifurcated optical fiber to obtain the circular-dichroic PL spectra. Furthermore, the cryostat is mounted on an automated translation stage, which enables circular-dichroic PL of large-area samples up to several hundred microns in size to be mapped efficiently. This technique allows non-destructive screening of CVD-grown WS$_2$ samples for homogeneity and batch-to-batch variations, for example by comparison of the linewidths and degrees of circular polarization.

"Environmental" and temperature-related effects can also impact of the charge transport properties of TMDC materials as indicated by the large variability of hole/electron mobility obtained even for a specific materials under same nominal experimental condition.[59,60] A detailed understanding on how defects, substrate, temperature and any other related fabrication and operational conditions affect the single layer optoelectronic properties is critical for controlling charge and spin transport in TMDC based device and therefore to encode any information in their valley states. Despite recent progress in the field,[61–63] a full picture of the above processes and their consequent impact is yet to be provided. Useful information on the electronic properties near the valley regions of single layer TMDCs can also be provided by angular resolved photoemission spectroscopy (ARPES), where a monochromatic radiation is incident on a sample and electrons are consequently emitted into the vacuum.[54,64] By analyzing the electron kinetic energy and angular distribution, this technique allows a direct determination of the electronic band structure along high symmetry directions of the Brillouin zone.[54]

ARPES studies on the electronic properties on TMDC layered materials typically involved small size (~μm) TMDC flakes obtained by direct exfoliation from single bulk crystal and then transferred onto a conductive substrate.[65–69] Band structure investigations were therefore mainly conducted by micro-ARPES techniques at synchrotron radiation facilities where the reduced photon beam size (~μm) combined with microscopy techniques can be achieved to allow an accurate selection of the probed μm-size flakes. This technique was successfully employed in mapping the electronic band dispersion of TMDCs exfoliated layer of various thicknesses revealing, for example, the change of the valence band maximum position in the reciprocal space from K to Γ point and the increase of band branches near Γ when the layer thickness increases from monolayer to multilayer (**Figure 4**a,b).[65–69]

Large scale single layer TMDCs (of up to ~100 μm of domain lateral size) were also grown by physical[70,71] and chemical deposition[72–75] processes on single crystal metallic substrate, the electrical conductivity of the substrate being required to avoid charging effects during photoemission in ARPES measurements. Via epitaxy, high quality and large crystal domains (of up to ~100 μm) can be achieved, thus allowing a detailed characterization by more conventional ARPES systems, where higher energy/momentum resolution limit can be generally achieved (**Figure 4**c).

Even though ARPES studies can yield important information on the interface-related effects on single layer TMDCs electronic properties, the critical parameters governing the interfacial potential (i.e. layer-substrate distance, defects, etc.) are normally difficult to control experimentally. Progress is made in a recent ARPES study on the $MoS_2$(single layer)/HOPG interface,[30,76] where the interfacial potential landscape was tuned by temperature-induced change in the layer-substrate separation. These changes are reflected in the slight distortion of the valence band dispersion, i.e. change in the energy difference $\Delta E_{\Gamma K}$ of the extrema (Γ and K) of the Brillouin zone (**Figure 5**a,b) pointing to the increased proximity of the $MoS_2$ layer to the HOPG substrate as temperature is decreased from 295 K to 11 K, in accord with dedicated band structure calculations as shown in **Figure 5**d. Note that the calculated trend in **Figure 5**c rules out in-plane contraction as the cause of the change in $\Delta E_{\Gamma K}$ since it has an opposite trend.

The impact of interface tuning was evidenced by significant changes observed in the line shape of energy distribution curves (EDCs) at Γ point with temperature (**Figure 5**e), where the out of plane localization of the electronic wave function makes the charge dynamics more sensitive to change in the interfacial potential landscape (**Figure 5**f). In contrast, the EDC line shape at the K point is basically unaffected by temperature change (**Figure 5**e), a result favored by the strong in plane localization of electronic state near the valley point (**Figure 5**f). The above results suggest that the charge and locked spin dynamics at the K valley is essentially protected from the local change in the interfacial potential landscape originating from the increased proximity to the substrate at lower temperatures. Such potential variations could arise, for example, from structural inhomogeneity in the TMDC/substrate interface resulting from the change in the layer substrate separation and/or presence of impurities. This demonstration of the "immunity", of the valley locked spins to interfacial landscape may have crucial implications for hole transport in single layer devices seeking to exploit valley pseudospins and the control thereof in TMDC based heterointerfaces.

By proper modifications of the electron detection scheme,[54] the ARPES technique can also be utilized to directly address the spin polarization of valley states in TMDCs. With respect to the electronic band studies, such Spin-resolved ARPES (SARPES) measurements of TMDC materials are comparatively less reported, mainly as a consequence of the lower detection efficiency of the SARPES technique.[77–81] While a local laboratory based SARPES is beneficial for expedient feedback on materials development, the typically lower photon intensity and energy resolution in such systems, compared to synchrotron SARPES, limits the range of application only to materials whose spin-splits bands are sufficiently far apart in energy.

The SARPES measurements on single layer of of MoSe$_2$ and WSe$_2$ on bilayer graphene substrate were reported by references [77,78] showing a low out-of-plane valley spin polarization value ($P_z$ = 0.1~0.2) with respect to the sample surface. The reduction of the measured spin polarization with respect to the theoretical value i.e. $P_z$ = 1, (corresponding to full out of plane spin polarization) was attributed to (i) limited energy resolution of spin detector and (ii) overlap between the signals of twin domains (relative rotation of 60º) by epitaxy on graphene. In particular, the the latter effect results in a superimposition of the spin signal from the K and K' valley points of different domains and thus reduce the measured spin polarization value.

Interestingly, a recent theoretical study suggested that the spin texture of the TMDCs could be probed by SARPES even in the inversion symmetric bulk crystals (termed "hidden" spin-polarization), as a result of the localization of two spin-degenerated valence band maxima on different layers of the unit cell.[82] The existence of spin-polarized electronic states was demonstrated by synchrotron based SARPES investigation on high quality 2H-WSe$_2$[79] and 2H-MoS$_2$ single crystal.[80]. With the availability of larger TMDC films grown by CVD or other techniques (e.g. Physical Vapour Deposition), laboratory-based SARPES system[83] become viable options for local laboratories to provide rapid feedback to the TMDC grower for the purpose of material optimization, instead of large synchrotron facilities. Using such a laboratory-based SARPES system,[83] the local ("hidden") spin polarizations in WS$_2$ bulk layers were detected for the first time in a local laboratory (**Figure 6**).[84]

With the availability of the above tools for assessing the spins localized in valleys, it becomes possible now to address important materials development questions such as whether substrate effect, dielectric capping, TMDC defects, and impurities could be detrimental to the preservation of spin-valley coupling in TMDCs. Bussolotti et al.[76] addressed the substrate impact and found the K-valley hole dynamics in TMDC monolayers to be well-protected, while Moody et al.[61] found that certain species of Se-vacancy defects induced in WSe$_2$ by electron irradiation actually formed defect-bound exciton states with longer lifetimes indicating enhanced spin-valley coupling. These studies point to the tolerance of TMDCs to likely substrate interactions and material defects which are unavoidable in practical fabrication scenarios.

## 3. TMDC-based Qubits

### 3.1. Theoretically Proposed Architectures

In semiconductor systems, it is possible to spatially confine charge carriers down to the few-electron regime in potential wells using electrostatic gates. The resulting devices, also known as quantum dots or single electron transistors, allow for individual control of single charge or spin.[85] In such systems, the spin states can form a qubit basis for quantum computation, as described in the original proposal by Loss and DiVincenzo in 1998.[86] The five criteria necessary for universal quantum computation are **(i)** a scalable physical system with well characterized qubits, **(ii)** qubit initialization, **(iii)** long decoherence times longer than gate operation times, **(iv)** two-qubit operation and **(v)** qubit readout. Many of these five criteria have been largely fulfilled in studies of solid state quantum dot systems such as GaAs, silicon, graphene, Si/SiGe, semiconductor nanowires and carbon nanotubes etc.[85]

Electrostatically defined quantum dots can also be created in 2D TMDCs. As a class of 2D semiconductors, single (or mono) and few layers of TMDCs offer several interesting properties advantageous for creating qubits. These atomically thin semiconductors offer natural carrier confinement in one spatial dimension. The monolayers possess a sizeable direct band gap of ~1.5 to 2 eV in the optical range allowing electrostatic confinement and optical manipulation of carriers.[87] TMDCs also have several isotopes with zero nuclear spin which minimizes hyperfine interactions with the electron spin. Isotopic purification in other material systems have already been proven to boost decoherence times which is advantageous for criteria (iii).[88] Furthermore, TMDCs possess an additional valley pseudospin that can be flexibly utilized as a valley qubit or be combined with the electron spin to form a spin-valley qubit. TMDCs also exhibit very strong spin-orbit coupling, in contrast to other spin-valley systems such as graphene and carbon nanotubes, and thus offer the potential for fast gate operations on TMDC qubits.[23,24]

Several theoretical studies have explored the viability of TMDC qubits, including valley qubits, spin qubits, spin-valley qubits and even impurity based qubits.[23–25,87,89–91] **Figure 7** shows the schematics of theoretical proposals for a TMDC based (a) valley qubit,[90] (b) spin-valley qubit[23] and (c) impurity-assisted qubit.[25] A key advantage that these schemes offer is the ability to manipulate the qubits by fully electrical means which could provide a potentially less complex route for scale-up.

For the valley qubit,[90] the design in **Figure 7**a uses gate electrodes G1 - G4 to create an external electrostatic potential in order to confine a single electron. The application of oscillatory voltages to opposite gate electrodes can modulate the electron confinement potential and in turn the electron wave function. These induce transitions of the electron between the valley states. To exploit the spin-valley coupling in TMDCs, Kormányos et al. proposed the spin-valley qubit architecture in **Figure 7**b, using DFT calculations to support their analyses for WS$_2$, WSe$_2$, MoSe$_2$ and WSe$_2$.[23] Their results show that in order to implement the spin-valley qubit in quantum dots the level spacing in the quantum dot should be larger than thermal energy. The mean level spacing in a quantum dot is given by: $\Delta E = \frac{2\pi\hbar^2}{m_{\text{eff}}A}$, where $A$ is the area of the dot and $m_{\text{eff}}$ is the effective mass. Based on this analysis, one expects relatively small quantum dots are necessary to ensure that energy level spacing in the quantum dot is larger than the thermal energy. For instance, in $MoS_2$ with an $m_{eff}$ =0.7$m_e$ and a quantum dot radius ($r$)=40 nm, $\Delta E$ ~150 µeV.[34] Similarly, for $WS_2$: $m_{eff}$ =0.35$m_e$, $r$=40 nm, and $\Delta E$ ~270µeV. For $WSe_2$ $m_{eff}$ =0.4 $m_e$, $r$=40 nm, and $\Delta E$ ~240 µeV. Such level spacings can be adequately resolved in transport spectroscopy performed in dilution refrigerator with a base temperature of 10 mK (typical electron temperature = 100 mK corresponding to $k_B T_e$ = 8.6 µeV, where $T_e$ = electron temperature). In this respect, TMDCs with smaller effective masses, such as $WS_2$ and $WSe_2$, might be more advantageous compared to $MoS_2$ and $MoSe_2$. However, for the same effective magnetic field, the splitting between different valley states is significant larger for $MoS_2$ compared to $WS_2$, suggesting that $MoX_2$ compounds are more suitable for spin and valley filtering. The authors conclude that the most realistic approach in terms of the choices for qubit states is to use the lowest Kramers pairs at around zero magnetic field as a spin-valley qubit. An external magnetic field can then be used to tune the energy splitting between these two states. This also means that the relaxation time of such spin-valley qubits will only be limited by the longer spin or valley relaxation time while the pure dephasing time will be limited by the shorter of the two. The exchange interaction can then be utilized to couple adjacent spin-valley qubits for operating two-qubit gates.

For the impurity-assisted qubit,[25] Széchenyi et al. show that a short-range impurity such as a vacancy, substitutional atom or adatom in a monolayer TMDC qubit can couple the basis states of the spin-valley qubit. This allows for resonant qubit control via electrically driven spin resonance with the help of an in-plane magnetic field. In the case of $MoS_2$, an S-type impurity, e.g. a sulfur vacancy, the qubit Rabi frequencies were estimated to be on the order of 10 – 100 MHz. They conclude that $MoS_2$ has the smallest spin-orbit splitting and is likely the material best suited for their qubit architecture.

## 3.2. Experimental Demonstrations of Electrostatically Gated Quantum Dots

Despite the availability of several architectural proposals, experimental realization has proven challenging due to issues such as contacting and gating monolayer TMDCs. Low material and interface quality and the lack of a robust contact strategy remain key hindrances for progress. Yet, encouraging efforts in the last few years have reported the fabrication of electrostatically gated TMDC quantum dots with varying degrees of device quality and tunability.[34–38,40] They include single quantum dots in monolayer $MoS_2$,[34] trilayer $MoS_2$,[40] quasi-2D (>7 layers) $WSe_2$, $WS_2$, and few-layer $InSe_2$,[35,36,38] and double quantum dots in monolayer $MoS_2$ and quasi-2D $MoS_2$.[34,37]

*3.2.1. Fully Electrostatically Gated Quantum Dots*

Since 2015, there has been several report towards the development of valleytronic devices[92] and spin-valley qubits based on TMDCs.[34–40,54,83] The strategy is to employ electrostatic gating to form quantum dots and realize a scalable TMDC based qubit platform, relying on the concepts espoused in the theoretical proposals of **Figure 7**. These reports demonstrate the ability to electrostatically gate a range of 2D TMDC materials to achieve carrier confinement. **Figure 8** shows the schematic architectures of several electrostatically gated 2D TMDC quantum dot devices reported in literature.

While these works demonstrate the viability to electrostatically confine carriers in 2D TMDCs given sufficiently high-quality materials and suitable contacts and dielectrics, there is still no consensus on the best device architecture and fabrication process. Even in the choice of contacts to TMDCs, there is as yet no established standards for ready implementation. For few-layer InSe quantum dot, gated graphene was the material of choice for contacts. For monolayer and trilayer $MoS_2$, gated graphene and Ti were utilized. For quasi-2D $WS_2$ and $WSe_2$, Pd contacts were employed. Proper ohmic contacts are key for studying quantum transport in these systems but are difficult to achieve due to the work function mismatch with most metals and Fermi level pinning effects. However, recent works are increasingly pointing to ultraclean interface between metallic contacts and TMDCs as the critical factor in obtaining high quality ohmic contacts at low temperatures.[93,94] The presence of adsorbed contaminants on TMDC surfaces can result in the creation of interface states leading to Fermi level pinning and large contact resistance. Ultraclean metal-TMDC interfaces can be achieved through ultrahigh vacuum metal evaporation or using hBN as an etch mask.[94,95] In the latter approach, the hBN sheets are typically mechanically exfoliated and stacked onto the TMDC material in an inert environment, before etch windows are opened for evaporation in the hBN through selective plasma etching. However, ultrahigh vacuum evaporators are not common laboratory equipment, and hBN etch masks requiring assembly of mechanically exfoliated heterostructures is labor intensive and can be limited in scalability. Another promising approach is the use of indium metal deposited at low evaporation rate.[93] The low evaporation rate reduces damage to the TMDC film via kinetic energy transfer between the metal atoms and the 2D TMDC. Cross-sectional annular dark-field scanning transmission electron microscopy reveal ultraclean interfaces between the In and TMDC film, while electrical measurements reveal record low contact resistances across a variety of TMDC films such as $MoS_2$, $WS_2$, $WSe_2$ and $NbS_2$. Consequently, the high-quality contacts formed by these different techniques based on ultraclean TMDC-contact interfaces allow for high carrier mobilities up to ~$10^4$ cm$^2$/Vs on exfoliated samples and ~$10^2$ on CVD samples.

Aside from contacts, suitable dielectrics for local confinement electrostatic gates are also important. In electrostatically gated quantum dots, disorder limits device mobility and homogeneity, which prevents precise control over the confinement potential shape and tunnel couplings. These problems were observed in the quasi-2D $MoS_2$, $WSe_2$ and $WS_2$ quantum dots devices demonstrated using $SiO_2$ substrate and $Al_2O_3$ high-k dielectrics.[36–38] In a typical low temperature stability measurements of quantum dot devices, diamond-shaped regions of suppressed conductance, also known as Coulomb diamonds, are signatures of single-electron transport through a quantum dot. Measurements of the quasi-2D $MoS_2$, $WSe_2$ and $WS_2$ devices reveal the existence of overlapping Coulomb diamonds, suggesting that more than a single quantum dot is formed. The tunnel couplings were also found to be dominated by impurity defined traps. As a result of the disordered potential, it was not possible to reach the few-electron regime. Such impurity dominated transport were also recently observed by Lau et al. in studies of dual-gated few-layer exfoliated $MoS_2$ and $WSe_2$ as shown in **Figure 9**.[39] While transport through the TMDC film could be independently pinched off by split top gates, the current shows only a single current step in contrast to the expected multiple regularly spaced steps seen in higher quality hBN encapsulated devices.[34,35] Our experiments on top-gated CVD monolayer $MoS_2$ films with $HfO_2$ dielectric likewise show strong disorder-defined tunnel couplings with Coulomb oscillations that are dominated by a single gate (**Figure 9**e,f), similar to observations by Wang et al. [40] In contrast, a quantum dot formed by the top gates over uniform 2DEG should display diagonal resonances in a $V_L$ versus $V_R$ plot, where $V_{L/R}$ refers to the top gate voltage applied to the left/right confinement gates.

Disorder was found to be greatly reduced in devices where the TMDC films were prepared in an inert environment and encapsulated with hBN to form hBN/TMDC/hBN van der Waals heterostructures. The devices made using such heterostructures include monolayer and trilayer $MoS_2$, few-layer $InSe_2$. In these devices, measurements reveal clear diagonal resonances in the $V_L$ versus $V_R$ plots, implying a well-defined quantum dot located at the center of the gate-defined confinement.

*3.2.2. Electrostatic Confinement in Nanoribbons and Nanotubes*

While full electrostatic gating could simplify process requirements, the total number of top-gates required to achieve effective confinement could potentially increase architectural complexity and limit scaling. In order to satisfy the condition on quantum dot size to resolve quantum dot level spacing, alternative ways of creating strong confinement with few top-gates may be pursued. One way to achieve this is by physically shrinking one dimension of the 2D host materials to form nanowires. This can be done in two ways. The first is a top-down approach where a large TMDC flake is etched into a one dimensional nanoribbon. The second approach relies on the bottom up growth of one dimensional TMDC nanotubes or nanoribbons by chemical vapour deposition or other synthesis methods. Thereafter, electrostatic gates across the wires are used to confine the quantum dots.

The top-down approach requires a fabrication process which minimizes any degradation in the performance of the nanoribbon based devices. Nanoribbons based devices have been extensively studied in graphene at room temperature[96] as well as at low temperature.[97,98] Graphene being a semimetal, has been shown to exhibit measurable currents down to few tens of nanometers even at cryogenic temperatures.[96,99–101] However, shrinking the dimensions while maintaining a measurable current is quite challenging in semiconducting TMDC. Firstly, the large Schottky barrier between metal and semiconducting TMDC which significantly suppresses device current at low temperatures. Secondly, the device current strongly depends on the channel dimensions. In single layer TMDC nanoribbons with widths below 100nm, maintaining measurable current through nanoribbons field effect transistor at low temperatures can be quite challenging. Fabrication of nanoribbon field effect transistor is at least a two step electron beam lithography (EBL) process (**Figure 10**a). For example, one method employs a first EBL step for ohmic contact definition and second EBL step to define a mask to etch out nanoribbon from a TMDC flake (**Figure 10**a).[31] Dry etching process have been developed to etch nanoribbons in TMDC using BCl$_3$,[102] SF$_6$ [31,103] and O$_2$ plasma.[104] Among the various processes developed, devices fabricated using SF$_6$ dry plasma etching process have resulted in excellent room temperature field effect transistors in single layer MoS$_2$ nanoribbons down to 50 nm size. Such transistors can exhibit high mobilities (up to 50 cm$^2$/Vs), steep sub-threshold slope (3.5 V/dec using 300 nm SiO$_2$ back gating), high current ON/OFF ratio (10$^5$) as well as high current density (38 µA/µm).[31]

A common feature observed in most reported nanoribbon field effect transistor experiments is that the mobility in nanoribbon devices were enhanced (**Figure 11**a, blue trace) compared to the mobility before etching the TMDC flakes (**Figure 11**a, red trace). Typical mobility measured in TMDCs flake field effect transistors are <10 cm$^2$/Vs. After etching the TMDC flake into a nanoribbon, the mobility increased up to 50 cm$^2$/Vs in single layer MoS$_2$.[31] The origin of enhanced mobility is not so well understood and remains the subject of further studies.

The bottom-up approach involves synthesis of one-dimensional TMDC nanotubes and nanoribbons. Growing single layer TMDC nanotube has been quite challenging however, and several experiments have shown synthesis of multiwall nanotubes with diameter ranging from 10 nm to several micrometers (**Figure 10b**).[105,106] With nanotubes grown vertically in well defined array, scaling of such devices might be possible as shown in other material systems.[107] Field effect transistor fabricated from multiwall MoS$_2$ nanotubes exhibit mobility of 43 cm$^2$/Vs, steep sub-threshold slope of 200 mV/dec, a current ON/OFF ratio of 10$^3$ as well as current density 1 µA/µm comparable to the etched nanoribbons.[106]

The requirement of physically shrinking the dimension of the host material down to few tens of nanometres as required to resolve the excitation spectrum in a quantum dot can be achieved with existing lithographic processes. However, the evaluation of their electrical properties at low temperature is crucial for implementing spin based qubits in nanoribbon and nanotubes based devices. **Figure 11**, reproduced from a recent report[108], shows the device characteristics of such a nanoribbon when cooled down from room temperature to low temperature (77 K). The authors report a shift in the threshold voltage with decrease in temperature as well as oscillatory current behavior at sufficiently low temperature (e.g. <4K). [104,108] The current oscillations observed are attributed to Coulomb blockade due to single electron tunnelling through a quantum dot as shown in inset of **Figure 11**b. The authors rule out other phenomenon such as conductance quantisation[109,110] and Fabry Perot[111,112] interference due to the very small mean free path (200nm) reported for encapsulated $MoS_2$.[113] These conclusions were further supported by the observation of diamond shaped domains in the 2D conductance (*dI/dV*) map as a function of source-drain and backgate gate voltage which are typical characteristics associated with Coulomb blockade (**Figure 11**b).

Unlike in nanoribbon devices, nanotubes devices with a direct comparison between room temperature and low temperature transport measurements have yet to be reported. However, Coulomb blockade in a multilayer $MoS_2$ nanotube has been demonstrated (**Figure 11**c).[105] Observed Coulomb blockade in both nanoribbon and nanotube devices are thus far due to accidental formation of quantum dots typically attributed to either external environment factors, intrinsic material system nuances, or combination of both. External environmental sources resulting in quantum dot originate from outside the material system and may include trap states (defects) in the $SiO_2$ substrate,[114,115] residues from the fabrication process,[116] and dirt on the TMDC material.[117] These sources result in charge localization in material system creating a quantum dot (**Figure 12**a) evidenced by Coulomb blockade at low temperature. Intrinsic sources resulting in quantum dot in both types of device may arise from the edge effects which includes microscopic roughness along the etched edges,[117] molecule bound to the edge,[118] or edge reconfiguration[119] (**Figure 12**b). In this regard, nanotube-based devices may offer advantage over nanoribbon-based devices since they do not suffer from the edge states allowing to eliminate accidentally formed quantum dot due to abrupt edges. However, one should take caution that while nanotubes may be scalable in synthesis, they may not easily be scalable for aligned incorporation onto a substance to enable large scale device processing. Hence, the use of nanotubes remains a controversial option despite some advantages.

Measures may be taken to minimize or eliminate the source of defects in nanoribbon and nanotube-based devices. Environmental defects may be largely reduced or eliminated by encapsulating the nanoribbon or nanotube in hBN. This could minimize not only the influence of substrate but also exposure to chemicals and solvents during the fabrication process. Intrinsic sources of defects may be mitigated by chemically functionalizing the channel to generate smooth edges especially in nanoribbon devices.[119] Alternatively, if ALD deposition of a suitable high-k dielectric can passivate the unwanted edges states, this would also facilitate using local gates to electrostatically define quantum dot as well as to tune out the unwanted states. An example of a proposed nanoribbon-based device combining the measures to minimize the environmental and intrinsic defects along with local gates is shown in **Figure 12**c. This approach may potentially lead to device with electrostatically controlled quantum dot in nanoribbon/nanotube based devices similar to other material platforms.[120,121]

Although the developments above demonstrate concerted efforts toward the promise of TMDC based qubits, 2D materials is a relatively recent field and TMDC based qubit development lags behind those based on more mature technological materials like Si and GaAs. Significant efforts are currently underway to demonstrate a tunable gate defined quantum dot is a key step towards fulfilling criteria **(i)**: a scalable physical system with well characterized qubits. To move beyond exfoliated TMDCs and van der Waals heterostructures which are limited in terms of scalability, new advances in material processing and device fabrication will be required in order to realize high quality scalable TMDC based qubits.

## 4. Summary, Challenges & Perspectives

The expectation of quantum computing is likely at an unprecedented high, but the hardware required to realize a universal programmable quantum computer has not yet arrived. Although quantum computers with few to tens of qubits are now commercially available on the cloud, the lagging hardware development signals that incumbent qubit technologies may be suffering from the scale-up bottleneck. Electrically controlled solid-state qubits, though faced with an initial high barrier of stringent materials engineering, have begun to come of age with the recent demonstration of few-qubits gates [6,16,122]. It is envisaged that such solid-state platforms have great scale-up potential and would be strong contenders for the universal quantum computer. In this progress report, the focus has been on a recent entrant into the solid-state qubit arena based on a relatively new class of 2D semiconductors known as transition metal dichalcogenides. The motivation of spin-valley coupling in 2D TMDCs for building robust qubits is explained, accompanied by brief descriptions of proposed architectures in the literature. The key purpose however is to update on the state-of-the-art in the development of such valley-based spin qubits and reveal remaining challenges associated with the essential research and engineering of the materials, interfaces, and device fabrication.

While qubits gates based on 2D TMDCs have not yet been demonstrated, significant progress and understanding have been achieved in building quantum dot devices in 2D TMDCs, and quantum confinement and Coulomb blockade manifestations have now been observed by a few groups including the authors herein. Such observations allude to single electron control and are important precursors toward optimized quantum dots which could serve as qubits. Challenges remain with regards to showing the full viability of such qubits and we provide a non-exhaustive mention of some of them. For materials engineering, there is an urgent need for reliable large area and high-quality 2D TMDCs, a reproducible ohmic contact strategy, an effective encapsulation of the 2D TMDC without destroying the desirable spin-valley coupling. An effective doping strategy, although less known or reported, is also likely required. From the quantum dot fabrication perspective, current lithographic techniques appear not to hinder the gate patterning required for quantum dots of a few tens of nanometres. However, it remains to be seen if the gating architectures, dielectric performance and control of crosstalk would be able to cope with the requirements for scale-up. These are near term challenges which are currently being addressed. In parallel, it is expected that developments for measuring the related coherence lifetimes and for qubit readout will follow as those would be critical determinants prior to qubit gate demonstrations. Although a latecomer to the solid-state qubit scene, the recent progress toward valley-coupled spins over a relative short period of time has been significant and encouraging, and this augurs well for the prospect of a solid-state spin-valley quantum computer.

**Acknowledgement**

This work was supported by the A*STAR 2D PHAROS Grant No. 1527000016 and A*STAR QTE Grant No. A1685b0005.

# References

[1] S. Boixo, S. V Isakov, V. N. Smelyanskiy, R. Babbush, N. Ding, Z. Jiang, M. J. Bremner, J. M. Martinis, H. Neven, *Nat. Phys.* **2018**, *14*, 595.

[2] A. W. Harrow, A. Montanaro, *Nature* **2017**, *549*, 203.

[3] R. Maurand, X. Jehl, D. Kotekar-Patil, A. Corna, H. Bohuslavskyi, R. Laviéville, L. Hutin, S. Barraud, M. Vinet, M. Sanquer, S. De Franceschi, *Nat. Commun.* **2016**, *7*, 13575.

[4] M. Veldhorst, H. G. J. Eenink, C. H. Yang, A. S. Dzurak, *Nat. Commun.* **2017**, *8*, 1766.

[5] M. Veldhorst, J. C. C. Hwang, C. H. Yang, A. W. Leenstra, B. De Ronde, J. P. Dehollain, J. T. Muhonen, F. E. Hudson, K. M. Itoh, A. Morello, A. S. Dzurak, *Nat. Nanotechnol.* **2014**, *9*, 981.

[6] M. Veldhorst, C. H. Yang, J. C. C. Hwang, W. Huang, J. P. Dehollain, J. T. Muhonen, S. Simmons, A. Laucht, F. E. Hudson, K. M. Itoh, A. Morello, A. S. Dzurak, *Nature* **2015**, *526*, 410.

[7] T. M. Roberson, A. G. White, *Quantum Sci. Technol.* **2019**, *4*, 020505.

[8] A. Acín, I. Bloch, H. Buhrman, T. Calarco, C. Eichler, J. Eisert, D. Esteve, N. Gisin, S. J. Glaser, F. Jelezko, S. Kuhr, M. Lewenstein, M. F. Riedel, P. O. Schmidt, R. Thew, A. Wallraff, I. Walmsley, F. K. Wilhelm, *New J. Phys.* **2018**, *20*, 080201.

[9] M. Riedel, M. Kovacs, P. Zoller, J. Mlynek, T. Calarco, *Quantum Sci. Technol.* **2019**, *4*, 020501.

[10] B. Sussman, P. Corkum, A. Blais, D. Cory, A. Damascelli, *Quantum Sci. Technol.* **2019**, *4*, 020503.

[11] M. G. Raymer, C. Monroe, *Quantum Sci. Technol.* **2019**, *4*, 020504.

[12] Y. Yamamoto, M. Sasaki, H. Takesue, *Quantum Sci. Technol.* **2019**, *4*, 020502.

[13] R. Srivastava, I. Choi, T. Cook, The Commercial Prospects for Quantum Computing **2016**, 1–48.

[14] B. E. Kane, *Nature* **1998**, *393*, 133.

[15] M. Y. Simmons, F. J. Ruess, K. E. J. Goh, T. Hallam, S. R. Schofield, L. Oberbeck, N. J. Curson, A. R. Hamilton, M. J. Butcher, R. G. Clark, T. C. G. Reusch, *Mol. Simul.* **2005**, *31*, 505.

[16] Y. He, S. K. Gorman, D. Keith, L. Kranz, J. G. Keizer, M. Y. Simmons, *Nature* **2019**, *571*, 371.

[17] D. Kotekar-Patil, A. Corna, R. Maurand, A. Crippa, A. Orlov, S. Barraud, L. Hutin, M. Vinet, X. Jehl, S. De Franceschi, M. Sanquer, *Phys. Status Solidi Basic Res.* **2017**, *254*, 1600581.

[18] H. Bohuslavskyi, D. Kotekar-Patil, R. Maurand, A. Corna, S. Barraud, L. Bourdet, L. Hutin, Y. M. Niquet, X. Jehl, S. De Franceschi, M. Vinet, M. Sanquer, *Appl. Phys. Lett.* **2016**, *109*, 193101.

[19] F. J. Rueß, W. Pok, T. C. G. Reusch, M. J. Butcher, K. E. J. Goh, L. Oberbeck, G. Scappucci, A. R. Hamilton, M. Y. Simmons, *Small* **2007**, *3*, 563.

[20] A. Laucht, R. Kalra, S. Simmons, J. P. Dehollain, J. T. Muhonen, F. A. Mohiyaddin, S. Freer, F. E. Hudson, K. M. Itoh, D. N. Jamieson, J. C. McCallum, A. S. Dzurak, A. Morello, *Nat. Nanotechnol.* **2017**, *12*, 61.

[21] A. Corna, L. Bourdet, R. Maurand, A. Crippa, D. Kotekar-Patil, H. Bohuslavskyi, R. Laviéville, L. Hutin, S. Barraud, X. Jehl, M. Vinet, S. De Franceschi, Y.-M. Niquet, M. Sanquer, *npj Quantum Inf.* **2018**, *4*, 1.

[22] Z. Gong, G.-B. Liu, H. Yu, D. Xiao, X. Cui, X. Xu, W. Yao, *Nat. Commun.* **2013**, *4*, 1.

[23] A. Kormányos, V. Zólyomi, N. D. Drummond, G. Burkard, *Phys. Rev. X* **2014**, *4*, 011034.

[24] A. David, G. Burkard, A. Kormányos, *2D Mater.* **2018**, *5*, 035031.

[25] G. Széchenyi, L. Chirolli, A. Pályi, *2D Mater.* **2018**, *5*, 035004.

[26] J. R. Schaibley, H. Yu, G. Clark, P. Rivera, J. S. Ross, K. L. Seyler, W. Yao, X. Xu, *Nat. Rev. Mater.* **2016**, *1*, 1.

[27] F. Bussolotti, H. Kawai, Z. E. Ooi, V. Chellappan, D. Thian, A. L. C. Pang, K. E. J. Goh, *Nano Futur.* **2018**, *2*, 032001.

[28] H. Yu, M. Liao, W. Zhao, G. Liu, X. J. Zhou, Z. Wei, X. Xu, K. Liu, Z. Hu, K. Deng, S. Zhou, J.-A. Shi, L. Gu, C. Shen, T. Zhang, L. Du, L. Xie, J. Zhu, W. Chen, R. Yang, D. Shi, G. Zhang, *ACS Nano* **2017**, *11*, 12001.

[29] Y. Cheng, K. Yao, Y. Yang, L. Li, Y. Yao, Q. Wang, X. Zhang, Y. Han, U. Schwingenschlögl, *RSC Adv.* **2013**, *3*, 17287.

[30] Y. F. Lim, K. Priyadarshi, F. Bussolotti, P. K. Gogoi, X. Cui, M. Yang, J. Pan, S. W. Tong, S. Wang, S. J. Pennycook, K. E. J. Goh, A. T. S. Wee, S. L. Wong, D. Chi, *ACS Nano* **2018**, *12*, 1339.

[31] D. Kotekar-Patil, J. Deng, S. L. Wong, C. S. Lau, K. E. J. Goh, *Appl. Phys. Lett.* **2019**, *114*.

[32] C. Huyghebaert, T. Schram, Q. Smets, T. Kumar Agarwal, D. Verreck, S. Brems, A. Phommahaxay, D. Chiappe, S. El Kazzi, C. Lockhart De La Rosa, G. Arutchelvan, D. Cott, J. Ludwig, A. Gaur, S. Sutar, A. Leonhardt, D. Marinov, D. Lin, M. Caymax, I. Asselberghs, G. Pourtois, I. P. Radu, *Tech. Dig. - Int. Electron Devices Meet. IEDM* **2019**, *2018-Decem*, 22.1.1.

[33] D. Akinwande, C. Huyghebaert, C. Wang, M. I. Serna, S. Goossens, L. Li, H.-S. Philip Wong, Frank H. L. Koppens, *Nature* **2019**, *573*, 507.

[34] R. Pisoni, Z. Lei, P. Back, M. Eich, H. Overweg, Y. Lee, K. Watanabe, T. Taniguchi, T. Ihn, K. Ensslin, *Appl. Phys. Lett.* **2018**, *112*, 123101.

[35] M. Hamer, E. Tóvári, M. Zhu, M. D. Thompson, A. Mayorov, J. Prance, Y. Lee, R. P. Haley, Z. R. Kudrynskyi, A. Patanè, D. Terry, Z. D. Kovalyuk, K. Ensslin, A. V. Kretinin, A. Geim, R. Gorbachev, *Nano Lett.* **2018**, *18*, 3950.

[36] X. X. Song, Z. Z. Zhang, J. You, D. Liu, H. O. Li, G. Cao, M. Xiao, G. P. Guo, *Sci. Rep.* **2015**, *5*, 16113.

[37] Z. Z. Zhang, X. X. Song, G. Luo, G. W. Deng, V. Mosallanejad, T. Taniguchi, K. Watanabe, H. O. Li, G. Cao, G. C. G. P. Guo, F. Nori, G. C. G. P. Guo, *Sci. Adv.* **2017**, *3*, 1.

[38] X.-X. X. Song, D. Liu, V. Mosallanejad, J. You, T.-Y. Y. Han, D.-T. T. Chen, H.-O. O. Li, G. Cao, M. Xiao, G.-C. C. G.-P. P. Guo, G.-C. C. G.-P. P. Guo, *Nanoscale* **2015**, *7*, 16867.

[39] C. S. Lau, J. Y. Chee, D. Thian, H. Kawai, J. Deng, S. L. Wong, Z. E. Ooi, Y.-F. Lim, K. E. J. Goh, *Sci. Rep.* **2019**, *9*, 8769.

[40] K. Wang, K. De Greve, L. A. Jauregui, A. Sushko, A. High, Y. Zhou, G. Scuri, T. Taniguchi, K. Watanabe, M. D. Lukin, H. Park, P. Kim, *Nat. Nanotechnol.* **2018**, *13*, 128.

[41] J. A. Wilson, A. D. Yoffe, *Adv. Phys.* **1969**, *18*, 193.

[42] Z. Cai, B. Liu, X. Zou, H. M. Cheng, *Chem. Rev.* **2018**, *118*, 6091.

[43] L. Tao, K. Chen, Z. Chen, W. Chen, X. Gui, H. Chen, X. Li, J. Bin Xu, *ACS Appl. Mater. Interfaces* **2017**, *9*, 12073.

[44] Y. L. Huang, Y. Chen, W. Zhang, S. Y. Quek, C. H. Chen, L. J. Li, W. T. Hsu, W. H. Chang, Y. J. Zheng, W. Chen, A. T. S. Wee, *Nat. Commun.* **2015**, *6*, 6298.

[45] S. Wu, C. Huang, G. Aivazian, J. S. Ross, D. H. Cobden, X. Xu, *ACS Nano* **2013**, *7*, 2768.

[46] X. Ling, Y. H. Lee, Y. Lin, W. Fang, L. Yu, M. S. Dresselhaus, J. Kong, *Nano Lett.* **2014**, *14*, 464.

[47] K. F. Mak, K. He, J. Shan, T. F. Heinz, *Nat. Nanotechnol.* **2012**, *7*, 494.

[48] H. Zeng, J. Dai, W. Yao, D. Xiao, X. Cui, *Nat. Nanotechnol.* **2012**, *7*, 490.

[49] D. Xiao, G. Bin Liu, W. Feng, X. Xu, W. Yao, *Phys. Rev. Lett.* **2012**, *108*, 196802.

[50] K. Hao, G. Moody, F. Wu, C. K. Dass, L. Xu, C. H. Chen, L. Sun, M. Y. Li, L. J. Li, A. H. MacDonald, X. Li, *Nat. Phys.* **2016**, *12*, 677.

[51] A. M. Jones, H. Yu, N. J. Ghimire, S. Wu, G. Aivazian, J. S. Ross, B. Zhao, J. Yan, D. G. Mandrus, D. Xiao, W. Yao, X. Xu, *Nat. Nanotechnol.* **2013**, *8*, 634.

[52] K. F. Mak, K. L. McGill, J. Park, P. L. McEuen, *Science* **2014**, *344*, 1489.

[53] J. Lee, K. F. Mak, J. Shan, *Nat. Nanotechnol.* **2016**, *11*, 421.

[54] F. Bussolotti, H. Kawai, Z. E. Ooi, V. Chellappan, D. Thian, A. L. C. Pang, K. E. J. Goh, *Nano Futur.* **2018**, *2*, 032001.

[55] Z. Wu, Z. Ni, *Nanophotonics* **2017**, *6*, 1219.

[56] Z. Lin, B. R. Carvalho, E. Kahn, R. Lv, R. Rao, H. Terrones, M. A. Pimenta, M. Terrones, *2D Mater.* **2016**, *3*, 022002.

[57] V. Chellappan, A. L. C. Pang, S. Sarkar, Z. E. Ooi, K. E. J. Goh, *Electron. Mater. Lett.* **2018**, 766.

[58] Z. E. Ooi, V. Chellappan, A. L. C. Pang, K. E. J. Goh, A cryogenic circular dichroic photoluminescence microscope **2019**, 164452.

[59] M. Amani, M. L. Chin, A. G. Birdwell, T. P. O'Regan, S. Najmaei, Z. Liu, P. M. Ajayan, J. Lou, M. Dubey, *Appl. Phys. Lett.* **2013**, *102*, 193107.

[60] Y. Liu, J. Guo, E. Zhu, L. Liao, S. J. Lee, M. Ding, I. Shakir, V. Gambin, Y. Huang, X. Duan, *Nature* **2018**, *557*, 696.

[61] G. Moody, K. Tran, X. Lu, T. Autry, J. M. Fraser, R. P. Mirin, L. Yang, X. Li, K. L. Silverman, *Phys. Rev. Lett.* **2018**, *121*, 57403.

[62] J. Kim, C. Jin, B. Chen, H. Cai, T. Zhao, P. Lee, S. Kahn, K. Watanabe, T. Taniguchi, S. Tongay, M. F. Crommie, F. Wang, *Sci. Adv.* **2017**, *3*, e1700518.

[63] L. Cheng, X. Wang, W. Yang, J. Chai, M. Yang, M. Chen, Y. Wu, X. Chen, D. Chi, K. E. J. Goh, J. X. Zhu, H. Sun, S. Wang, J. C. W. Song, M. Battiato, H. Yang, E. E. M. Chia, *Nat. Phys.* **2019**, *15*.

[64] A. Damascelli, Z. Hussain, Z. X. Shen, *Rev. Mod. Phys.* **2003**, *75*, 473.

[65] W. Jin, P. C. Yeh, N. Zaki, D. Zhang, J. T. Sadowski, A. Al-Mahboob, A. M. Van Der Zande, D. A. Chenet, J. I. Dadap, I. P. Herman, P. Sutter, J. Hone, R. M. Osgood, *Phys. Rev. Lett.* **2013**, *111*, 106801.

[66] W. Jin, P. C. Yeh, N. Zaki, D. Zhang, J. T. Liou, J. T. Sadowski, A. Barinov, M. Yablonskikh, J. I. Dadap, P. Sutter, I. P. Herman, R. M. Osgood, *Phys. Rev. B* **2015**, *91*, 121409(R).

[67] H. Yuan, Z. Liu, G. Xu, B. Zhou, S. Wu, D. Dumcenco, K. Yan, Y. Zhang, S. K. Mo, P. Dudin, V. Kandyba, M. Yablonskikh, A. Barinov, Z. Shen, S. Zhang, Y. Huang, X. Xu, Z. Hussain, H. Y. Hwang, Y. Cui, Y. Chen, *Nano Lett.* **2016**, *16*, 4738.

[68] C. Kastl, R. J. Koch, C. T. Chen, J. Eichhorn, S. Ulstrup, A. Bostwick, C. Jozwiak, T. R. Kuykendall, N. J. Borys, F. M. Toma, S. Aloni, A. Weber-Bargioni, E. Rotenberg, A. M. Schwartzberg, *ACS Nano* **2019**, *13*, 1284.

[69] S. Ulstrup, J. Katoch, R. J. Koch, D. Schwarz, S. Singh, K. M. McCreary, H. K. Yoo, J. Xu, B. T. Jonker, R. K. Kawakami, A. Bostwick, E. Rotenberg, C. Jozwiak, *ACS Nano* **2016**, *10*, 10058.

[70] M. Dendzik, M. Michiardi, C. Sanders, M. Bianchi, J. A. Miwa, S. S. Grønborg, J. V. Lauritsen, A. Bruix, B. Hammer, P. Hofmann, *Phys. Rev. B* **2015**, *92*, 245442.

[71] J. A. Miwa, S. Ulstrup, S. G. Sørensen, M. Dendzik, A. G. Čabo, M. Bianchi, J. V. Lauritsen, P. Hofmann, *Phys. Rev. Lett.* **2015**, *114*, 046802.

[72] Y. Zhang, T. R. Chang, B. Zhou, Y. T. Cui, H. Yan, Z. Liu, F. Schmitt, J. Lee, R. Moore, Y. Chen, H. Lin, H. T. Jeng, S. K. Mo, Z. Hussain, A. Bansil, Z. X. Shen, *Nat. Nanotechnol.* **2014**, *9*, 111.

[73] D. J. Trainer, A. V. Putilov, C. Di Giorgio, T. Saari, B. Wang, M. Wolak, R. U. Chandrasena, C. Lane, T. R. Chang, H. T. Jeng, H. Lin, F. Kronast, A. X. Gray, X. X. Xi, J. Nieminen, A. Bansil, M. Iavarone, *Sci. Rep.* **2017**, *7*, 40599.

[74] N. Alidoust, G. Bian, S. Y. Xu, R. Sankar, M. Neupane, C. Liu, I. Belopolski, D. X. Qu, J. D. Denlinger, F. C. Chou, M. Z. Hasan, *Nat. Commun.* **2014**, *5*, 4673.

[75] Y. Zhang, M. M. Ugeda, C. Jin, S.-F. Shi, A. J. Bradley, A. Martín-Recio, H. Ryu, J. Kim, S. Tang, Y. Kim, B. Zhou, C. Hwang, Y. Chen, F. Wang, M. F. Crommie, Z. Hussain, Z.-X. Shen, S.-K. Mo, *Nano Lett.* **2016**, *16*, 2485.

[76] F. Bussolotti, H. Kawai, S. L. Wong, K. E. J. Goh, *Phys. Rev. B* **2019**, *99*, 045134.

[77] K. Sugawara, T. Sato, Y. Tanaka, S. Souma, T. Takahashi, *Appl. Phys. Lett.* **2015**, *107*, 071601.

[78] S.-K. Mo, C. Hwang, Y. Zhang, M. Fanciulli, S. Muff, J. Hugo Dil, Z.-X. Shen, Z. Hussain, *J. Phys. Condens. Matter* **2016**, *28*, 454001.

[79] J. M. Riley, F. Mazzola, M. Dendzik, M. Michiardi, T. Takayama, L. Bawden, C. Granerd, M. Leandersson, T. Balasubramanian, M. Hoesch, T. K. Kim, H. Takagi, W. Meevasana, P. Hofmann, M. S. Bahramy, J. W. Wells, P. D. C. King, *Nat. Phys.* **2014**, *10*, 835.

[80] M. Gehlmann, I. Aguilera, G. Bihlmayer, E. Młyńczak, M. Eschbach, S. Döring, P. Gospodarič, S. Cramm, B. Kardynał, L. Plucinski, S. Blügel, C. M. Schneider, *Sci. Rep.* **2016**, *6*, 26197.

[81] E. Razzoli, T. Jaouen, M.-L. L. Mottas, B. Hildebrand, G. Monney, A. Pisoni, S. Muff, M. Fanciulli, N. C. C. C. Plumb, V. A. A. A. Rogalev, V. N. N. N. Strocov, J. Mesot,

M. Shi, J. H. H. H. Dil, H. Beck, P. Aebi, *Phys. Rev. Lett.* **2017**, *118*, 086402.

[82] X. Zhang, Q. Liu, J. W. Luo, A. J. Freeman, A. Zunger, *Nat. Phys.* **2014**, *10*, 387.

[83] F. Bussolotti, Z. Zhang, H. Kawai, K. E. J. Goh, In *MRS Advances*; 2017; Vol. 2, pp. 1527–32.

[84] *2D semiconductr materials and devices*; Chi, D.; Goh, K. E. J.; Wee, A. T. S., Eds.; Elsevier, 2019.

[85] R. Hanson, J. R. Petta, S. Tarucha, L. M. K. Vandersypen, L. P. Kouwenhoven, J. R. Petta, S. Tarucha, L. M. K. Vandersypen, *Rev. Mod. Phys.* **2007**, *79*, 1217.

[86] D. Loss, D. P. DiVincenzo, *Phys Rev A* **1998**, *57*, 120.

[87] Y. Wu, Q. Tong, G. Bin Liu, H. Yu, W. Yao, *Phys. Rev. B* **2016**, *93*, 043313.

[88] K. M. Itoh, H. Watanabe, *MRS Commun.* **2014**, *4*, 143.

[89] A. J. Pearce, G. Burkard, *2D Mater.* **2017**, *4*, 025114.

[90] J. Pawłowski, D. Zebrowski, S. Bednarek, *Phys. Rev. B* **2018**, *97*, 155412.

[91] M. Brooks, G. Burkard, *Phys. Rev. B* **2017**, *95*, 245411.

[92] J. Lee, Z. Wang, H. Xie, K. F. Mak, J. Shan, *Nat. Mater.* **2017**, *16*, 887.

[93] Y. Wang, J. C. Kim, R. J. Wu, J. Martinez, X. Song, J. Yang, F. Zhao, A. Mkhoyan, H. Y. Jeong, M. Chhowalla, *Nature* **2019**, *568*, 70.

[94] S. Xu, Z. Wu, H. Lu, Y. Han, G. Long, X. Chen, T. Han, W. Ye, Y. Wu, J. Lin, J. Shen, Y. Cai, Y. He, F. Zhang, R. Lortz, C. Cheng, N. Wang, *2D Mater* **2016**, *3*, 021007.

[95] C. D. English, G. Shine, V. E. Dorgan, K. C. Saraswat, E. Pop, *Nano Lett.* **2016**, *16*, 3824.

[96] W. S. Hwang, P. Zhao, K. Tahy, L. O. Nyakiti, V. D. Wheeler, R. L. Myers-Ward, C. R. Eddy, D. K. Gaskill, J. A. Robinson, W. Haensch, H. Xing, A. Seabaugh, D. Jena, *APL Mater.* **2015**, *3*, 011101.

[97] X. L. Liu, D. Hug, L. M. K. Vandersypen, *Nano Lett.* **2010**, *10*, 1623.

[98] Y. Song, H. Xiong, W. Jiang, H. Zhang, X. Xue, C. Ma, Y. Ma, L. Sun, H. Wang, L. Duan, *Nano Lett.* **2016**, *16*, 6245.

[99] S. Dufferwiel, T. P. Lyons, D. D. Solnyshkov, A. A. P. Trichet, F. Withers, S. Schwarz, G. Malpuech, J. M. Smith, K. S. Novoselov, M. S. Skolnick, D. N. Krizhanovskii, A. I. Tartakovskii, *Nat. Photonics* **2017**, *11*, 497.

[100] J. Bai, R. Cheng, F. Xiu, L. Liao, M. Wang, A. Shailos, K. L. Wang, Y. Huang, X. Duan, *Nat. Nanotechnol.* **2010**, *5*, 655.

[101] J. Sun, T. Iwasaki, M. Muruganathan, H. Mizuta, *Appl. Phys. Lett.* **2015**, *106*, 033509.

[102] H. Liu, J. Gu, P. D. Ye, *IEEE Electron Device Lett.* **2012**, *33*, 1273.

[103] F. Zhang, C. H. Lee, J. A. Robinson, J. Appenzeller, *Nano Res.* **2018**, *11*, 1768.

[104] Y. Li, N. Mason, *arXiv* **2013**, 1312.3939.

[105] S. Reinhardt, L. Pirker, C. Bäuml, M. Remškar, A. K. Hüttel, *Phys. status solidi – Rapid Res. Lett.* **2019**, *1900251*, 4.

[106] S. Fathipour, M. Remskar, A. Varlec, A. Ajoy, R. Yan, S. Vishwanath, S. Rouvimov, W. S. Hwang, H. G. Xing, D. Jena, A. Seabaugh, *Appl. Phys. Lett.* **2015**, *106*, 022114.

[107] G. Larrieu, X.-L. Han, *Nanoscale* **2013**, *5*, 2437.

[108] D. Kotekar-Patil, J. Deng, S. L. Wong, K. E. J. Goh, *ACS Appl. Electron. Mater.* **2019**, *in press*.

[109] M. J. Biercuk, N. Mason, J. Martin, A. Yacoby, C. M. Marcus, *Phys. Rev. Lett.* **2005**, *94*, 026801.

[110] E. Tóvári, P. Makk, M. H. Liu, P. Rickhaus, Z. Kovács-Krausz, K. Richter, C. Schönenberger, S. Csonka, *Nanoscale* **2016**, *8*, 19910.

[111] D. Kotekar-Patil, B. M. Nguyen, J. Yoo, S. A. Dayeh, S. M. Frolov, *Nanotechnology* **2017**, *28*, 385204.

[112] W. Liang, M. Bockrath, D. Bozovic, J. H. Hafner, M. Tinkham, H. Park, *Nature* **2001**, *411*, 665.

[113] R. Pisoni, Y. Lee, H. Overweg, M. Eich, P. Simonet, K. Watanabe, T. Taniguchi, R. Gorbachev, T. Ihn, K. Ensslin, *Nano Lett.* **2017**, *17*, 5008.

[114] J. Xue, J. Sanchez-Yamagishi, D. Bulmash, P. Jacquod, A. Deshpande, K. Watanabe, T. Taniguchi, P. Jarillo-Herrero, B. J. Leroy, *Nat. Mater.* **2011**, *10*, 282.

[115] S. Ghatak, A. N. Pal, A. Ghosh, *ACS Nano* **2011**, *5*, 7707.

[116] J. H. Chen, W. G. Cullen, C. Jang, M. S. Fuhrer, E. D. Williams, *Phys. Rev. Lett.* **2009**, *102*, 236805.

[117] D. Bischoff, A. Varlet, P. Simonet, M. Eich, H. C. Overweg, T. Ihn, K. Ensslin, *Appl. Phys. Rev.* **2015**, *2*, 031301.

[118] T. Kato, L. Jiao, X. Wang, H. Wang, X. Li, L. Zhang, R. Hatakeyama, H. Dai, *Small* **2011**, *7*, 574.

[119] X. Jia, M. Hofmann, V. Meunier, B. G. Sumpter, J. Campos-Delgado, J. M. Romo-Herrera, H. Son, Y. P. Hsieh, A. Reina, J. Kong, M. Terrones, M. S. Dresselhaus, *Science (80-. ).* **2009**, *323*, 1701.

[120] X. Jehl, B. Roche, M. Sanquer, B. Voisin, R. Wacquez, V. Deshpande, B. Previtali, M. Vinet, J. Verduijn, G. C. Tettamanzi, S. Rogge, D. Kotekar-Patil, M. Ruoff, D. Kern, D. A. Wharam, M. Belli, E. Prati, M. Fanciulli, *Procedia Comput. Sci.* **2011**, *7*, 266.

[121] S. Nadj-Perge, S. M. Frolov, E. P. A. M. Bakkers, L. P. Kouwenhoven, *Nature* **2010**, *468*, 1084.

[122] T. F. Watson, S. G. J. Philips, E. Kawakami, D. R. Ward, P. Scarlino, M. Veldhorst, D. E. Savage, M. G. Lagally, M. Friesen, S. N. Coppersmith, M. A. Eriksson, L. M. K. Vandersypen, .

**Figure and Captions**

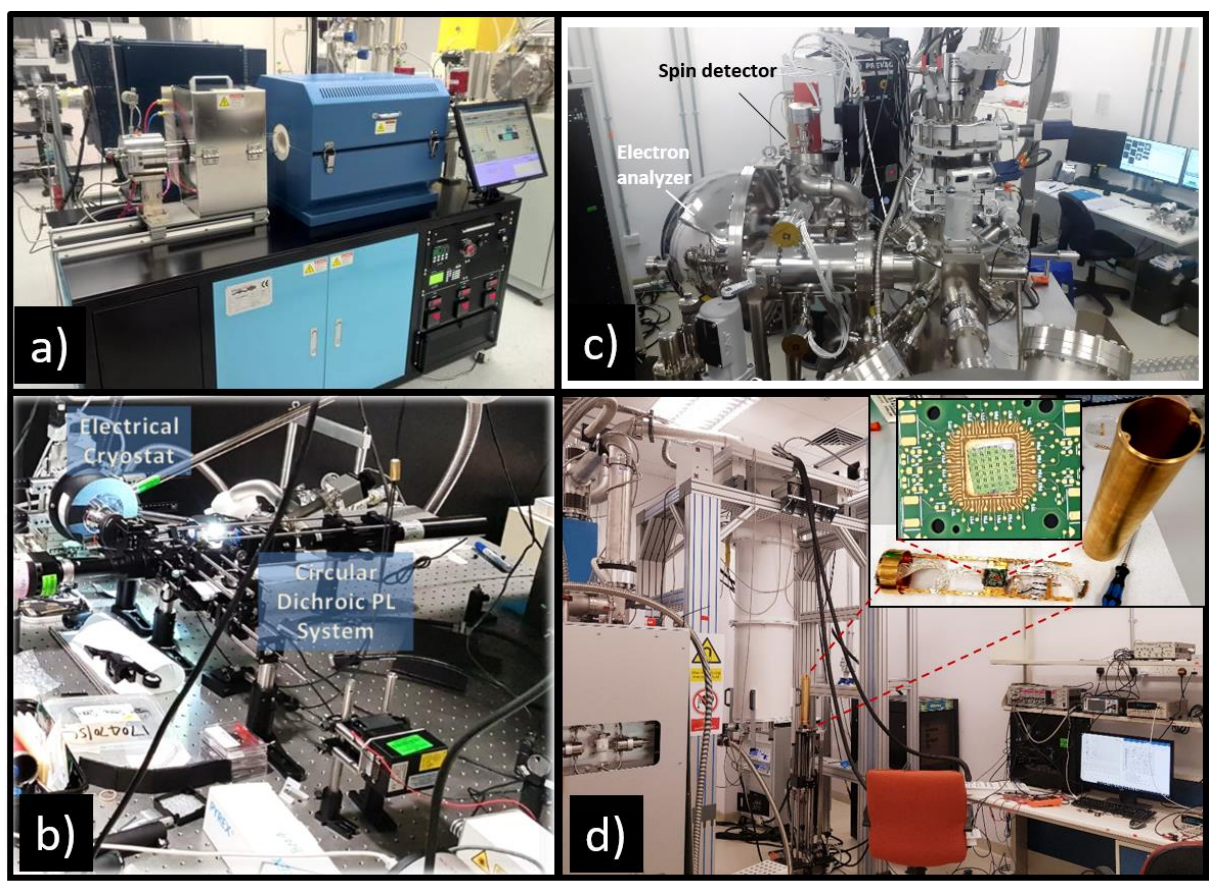

**Figure 1.** A snap shot of the key facilities available at A*STAR (Singapore) for the growth and characterization of transition metal dichalcogenides (TMDC) materials.[27,83] The system shown in (a) is a 3 zone furnace for growing high quality TMDC layer by Chemical Vapour Deposition (CVD) technique. The quality of as-grown TMDC materials can be typically assessed by optical microscopy and Raman spectroscopy, but the Circular Dichroic Photoluminescence spectroscopy(CDPL) set-up in (b) allows a rapid, non-invasive screening of valley polarization in two dimensional materials. TMDC's electronic band structure, which ultimately determine the electrical and optical response of the layers, as well as their valley and spin polarization can also be independently detected by Spin and Angular Resolved Photoelectron spectroscopy (SARPES) in (c). CDPL and SARPES represent critical experimental tools to quantitatively rationalize the impact of supporting substrate and/or structural and chemical defect introduced by the sample preparation process on the electronic and optical properties of the TMDC layers utilized in the fabricated quantum devices. These devices can be investigated in dilution refrigerators enabling electrical transport measurements at temperatures down to 10 mK (d).

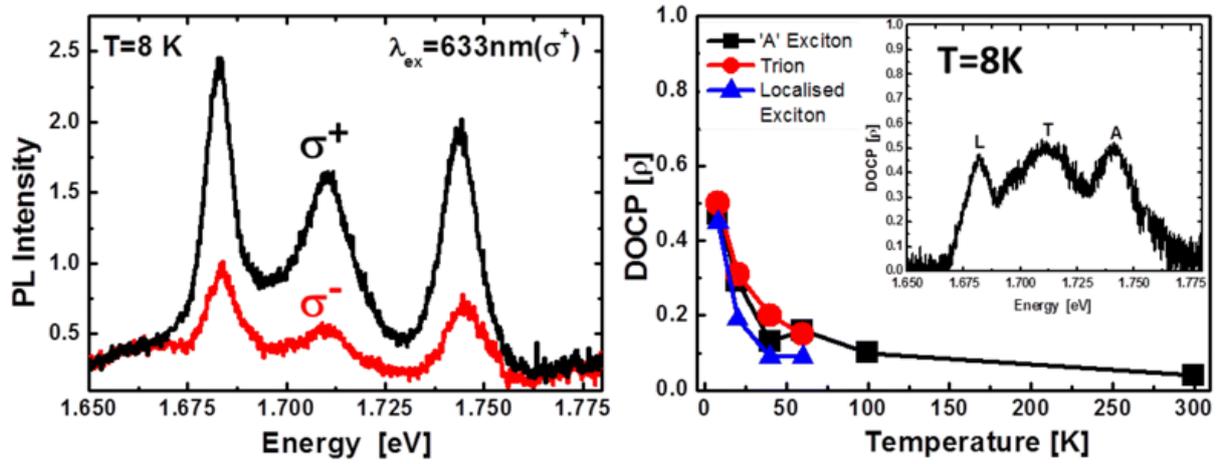

**Figure 2.** Circular Dichroic Photoluminesence (CDPL) spectroscopy of WSe$_2$. *Left*: CDPL response of a WSe$_2$ single layer at 8 K. *Right*: Valley polarization inferred from the Degree of Circular Polarization (DOCP) extracted from the CDPL intensity for various optical excitation (Exciton, Trions, localized exciton). Inset shows the DOCP at a temperature of 8 K. Reproduced with permission. [57] 2018, Springer Nature.

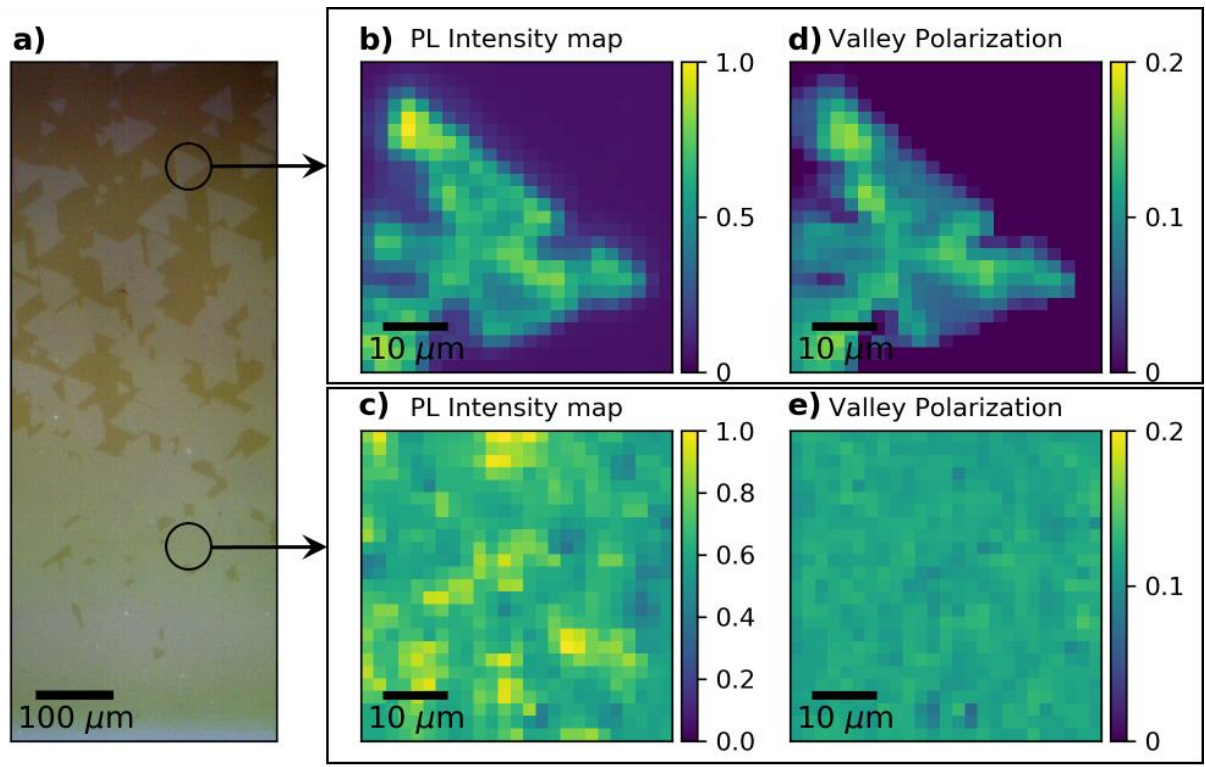

**Figure 3.** (a) Optical image of Chemical Vapor Deposition (CVD) grown single-layer $WS_2$ on single-crystal sapphire. The $WS_2$ is grown at 760 torr in an $Ar/H_2$ atmosphere at a maximum temperature of 850 $^oC$. (b,c) Photoluminescence (PL) maps of (b) an isolated single crystal grain and (c) the continuous single-layer film, measured at 4K in an optical cryostat. (d,e) Valley polarization of the respective regions in (b, c). The mapping of the valley-polarized PL enables non-destructive screening of CVD-grown $WS_2$ samples to determine sample homogeneity and quality (especially the degree of valley selectivity), before device fabrication.

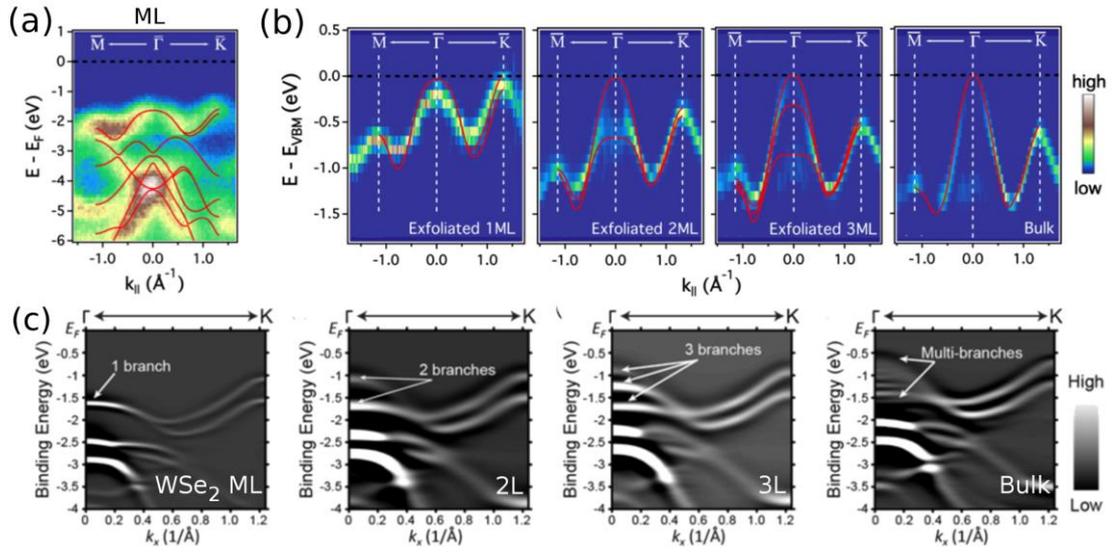

**Figure 4**. ARPES studies on TMDCs. (a) Micro-ARPES spectra from exfoliated single layer MoS$_2$. (b) Experimental band dispersion along the MΓK high symmetry directions of exfoliated monolayer, bilayer, trilayer and multilayer (bulk-like) of MoS$_2$ (right panels). The electronic band dispersions were obtained by second derivative filter of ARPES data (not shown). Figures in panel (a) and (b) adapted from Ref. [65]. (c) Experimental band dispersion along the ΓK high symmetry direction of monolayer, bilayer, trilayer and multilayer (bulk-like) WSe$_2$ on bilayer graphene, as obtained by second derivative filtering of corresponding ARPES data (not shown). Reproduced with permission.[75] 2016, American Chemical Society.

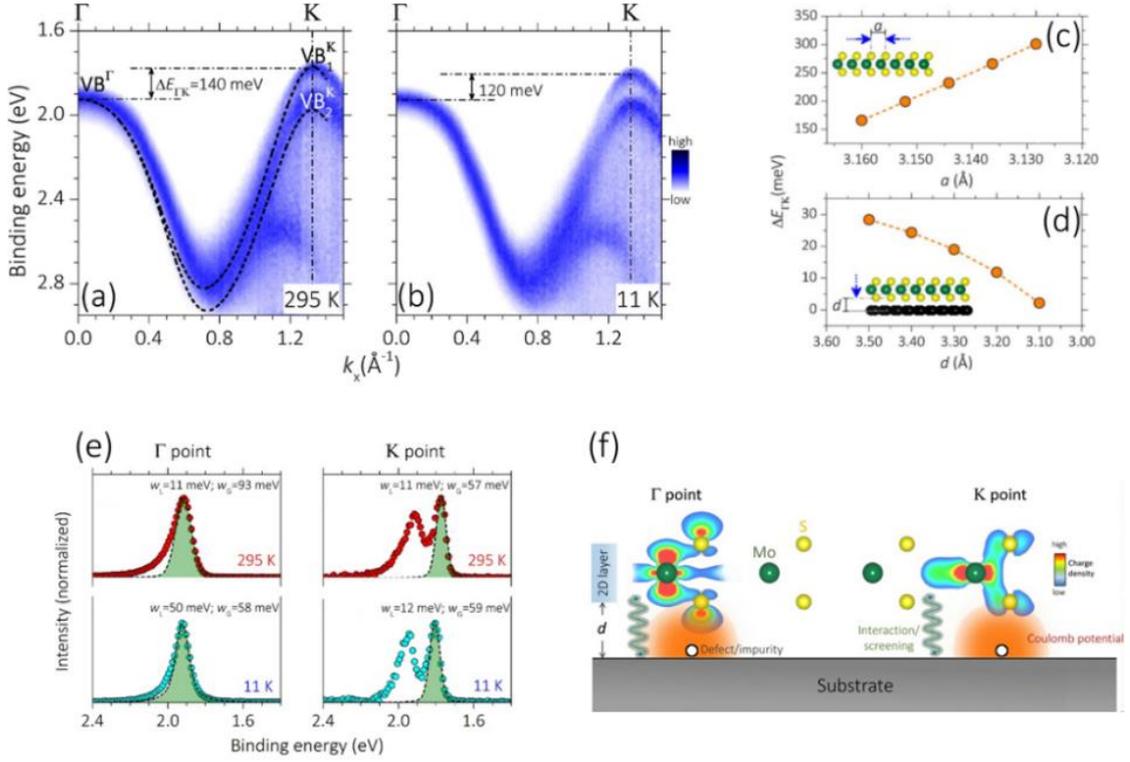

**Figure 5.** Angular Resolved Photoemission Spectroscopy (ARPES) study of the single layer $MoS_2$ on HOPG substrate. (a,b) ARPES intensity plot (normalized to maximum) showing the VB dispersion of single layer $MoS_2$ along ΓK at 295 K (a) and 11 K (b). Dashed black curves in panel (a) denotes the calculated band dispersion for free-standing single layer $MoS_2$. The energy separation between the band extrema at Γ and K point ($\Delta E_{\Gamma K}$) is also indicated at both temperatures. (c,d) Calculated $\Delta E_{\Gamma K}$ with varying lattice constant $a$ in a free-standing single layer (c) and distance $d$ from supporting substrate (single layer graphite used, for computational efficiency). (e) Energy Distribution Curves (EDCs) at Γ for 295 K and 11 K (left panels) and at K point for 295 K and 11 K (right panels). EDCs near VB local maximum were fitted by Voigt functions, (dashed curve with green shading). The extracted Lorentzian ($w_L$) and Gaussian ($w_G$) are reported in panels each panel. (f) Schematic impact of substrate interaction and impurities/defects Coulomb potential on the single layer $MoS_2$ electronic states near the local VB maximum at Γ and K. Calculated charge density plot reflects the wavefunction symmetry at different points of the Brillouin zone. Reproduced with permission.[76] 2019, American Physical Society.

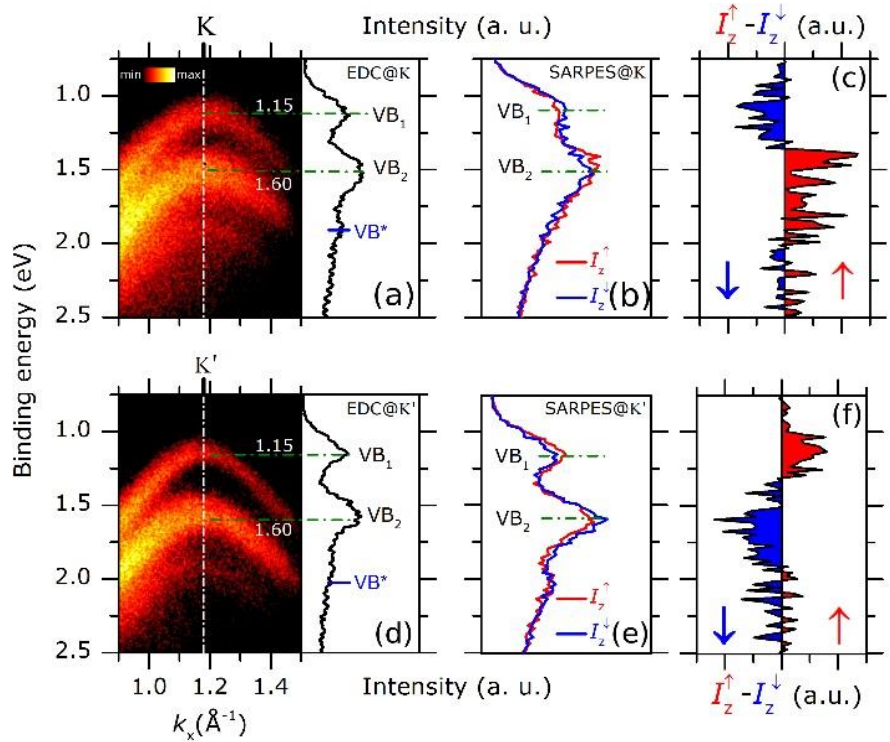

**Figure 6**. Laboratory based Spin Angular Resolved Photoemission Spectroscopy (SARPES) measurements of 2H-WS2 bulk single crystal. (a) ARPES intensity of WS$_2$ bulk single crystal as a function of binding energy and momentum ($k_x$) as acquired around K point of a WS$_2$ single crystal (left) and corresponding energy distribution curves (right). (b,c) SARPES signal at K valley point of the WS$_2$ Brillouin zone (b), as measured along the *z*-direction of the spin detector ($I_z^\uparrow$ and $I_z^\downarrow$), and corresponding signal difference(c). (d-f) Same as panels (a)-(c) for K' point. The inversion of signal difference in panel (c) and (f) corresponds to opposite spin polarization at K and K' valley. Reproduced with permission.[84] 2019, Elsevier.

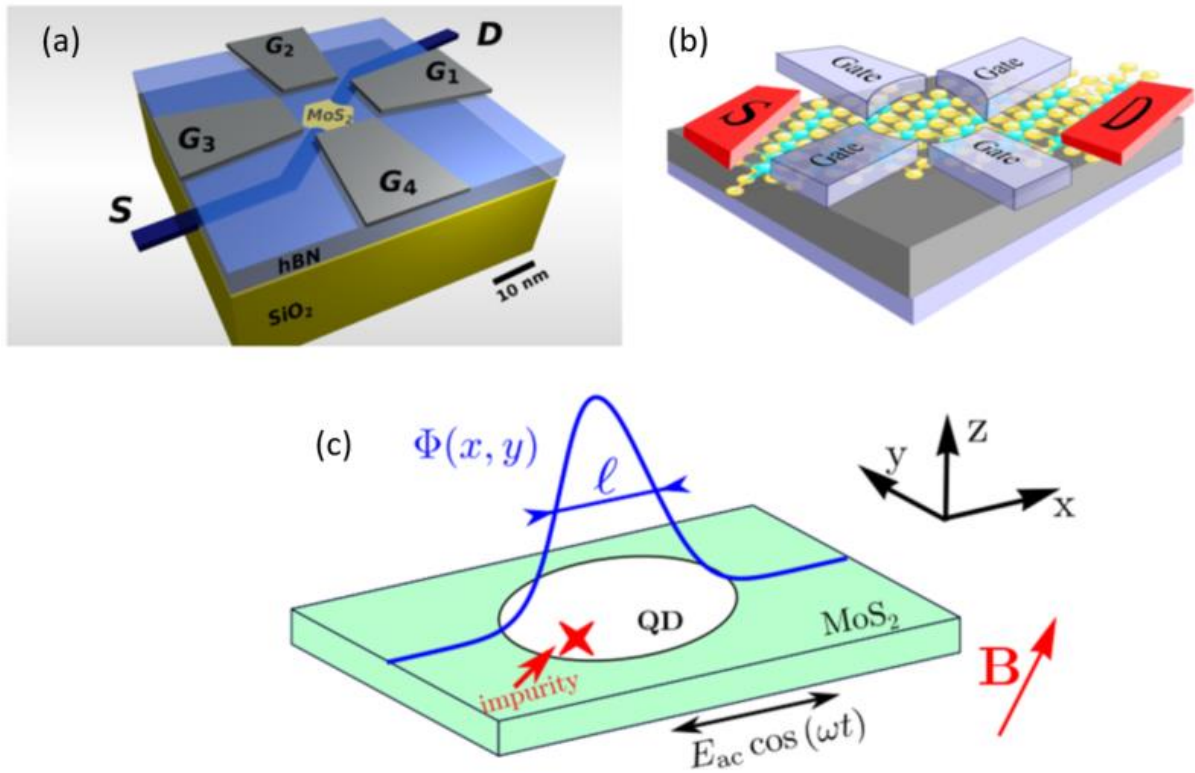

**Figure 7.** Schematics of proposed qubit architectures based on quantum dots confined in 2D TMDCs. a) Valley qubit.[90] b) Spin-valley qubit.[23] c) Impurity-assisted qubit.[25] Reproduced with permission.[23,25,90] 2014, American Physical Society. 2018, Institute of Physics. 2018, American Physical Society.

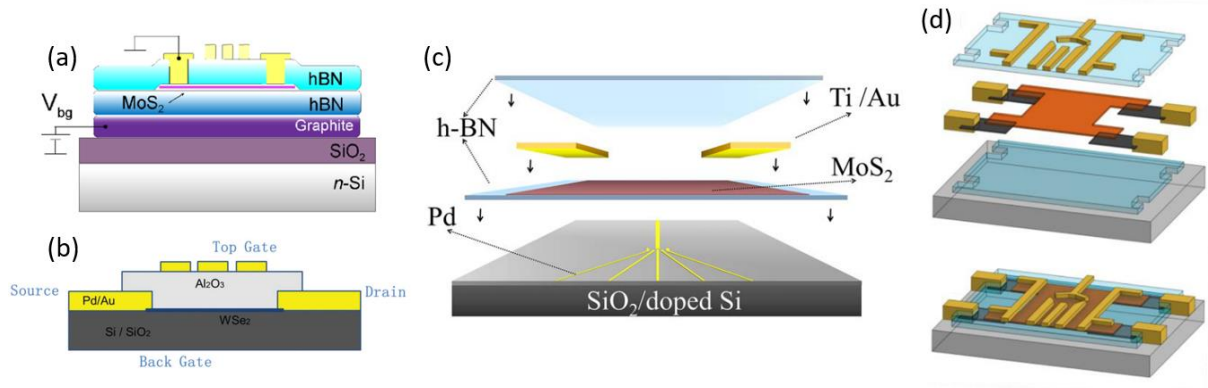

**Figure 8.** Schematics of different electrostatically gated transition metal dichalcogenide quantum dot devices reported in literature. (a) An exfoliated single layer $MoS_2$ device encapsulated between two hBN layers.[34] An electron mobility as high as 3000 $cm^2$/Vs was estimated from the onset of Shubnikov-de Haas oscillations. By tuning the local confinement gate voltages, electrons could be confined in single or double dot regimes depending on the voltage applied on the middle gate. Reproduced with permission.[34] 2018, AIP Publishers. (b) A seven-layer thick exfoliated $WSe_2$ device capped with atomic layer deposited $Al_2O_3$. The device exhibited clear Coulomb diamonds when modulating either the plunger gate or the back gate, indicating single electron transport. Reproduced with permission.[38] 2015, Royal Society of Chemistry. (c) An exfoliated eight-layer thick $MoS_2$ device encapsulated between two hBN layers. The field effect mobility was estimated to be ~300 $cm^2$/Vs. The device could also be tuned between a single and double dot regime depending on the confinement gate voltages. Reproduced with permission.[37] 2017, AAAS. (d) A six-layer thick exfoliated $InSe_2$ device encapsulated between two hBN layers. A mobility on the order of 10,000 $cm^2$/Vs was estimated from the onset of Shubnikov-de Haas oscillations. Conductance quantization with multiple evenly spaced conductance steps was observed with the InSe point contacts. Coulomb diamonds were also observed when modulating the local confinement gates. Reproduced with permission.[35] 2018, American Chemical Society.

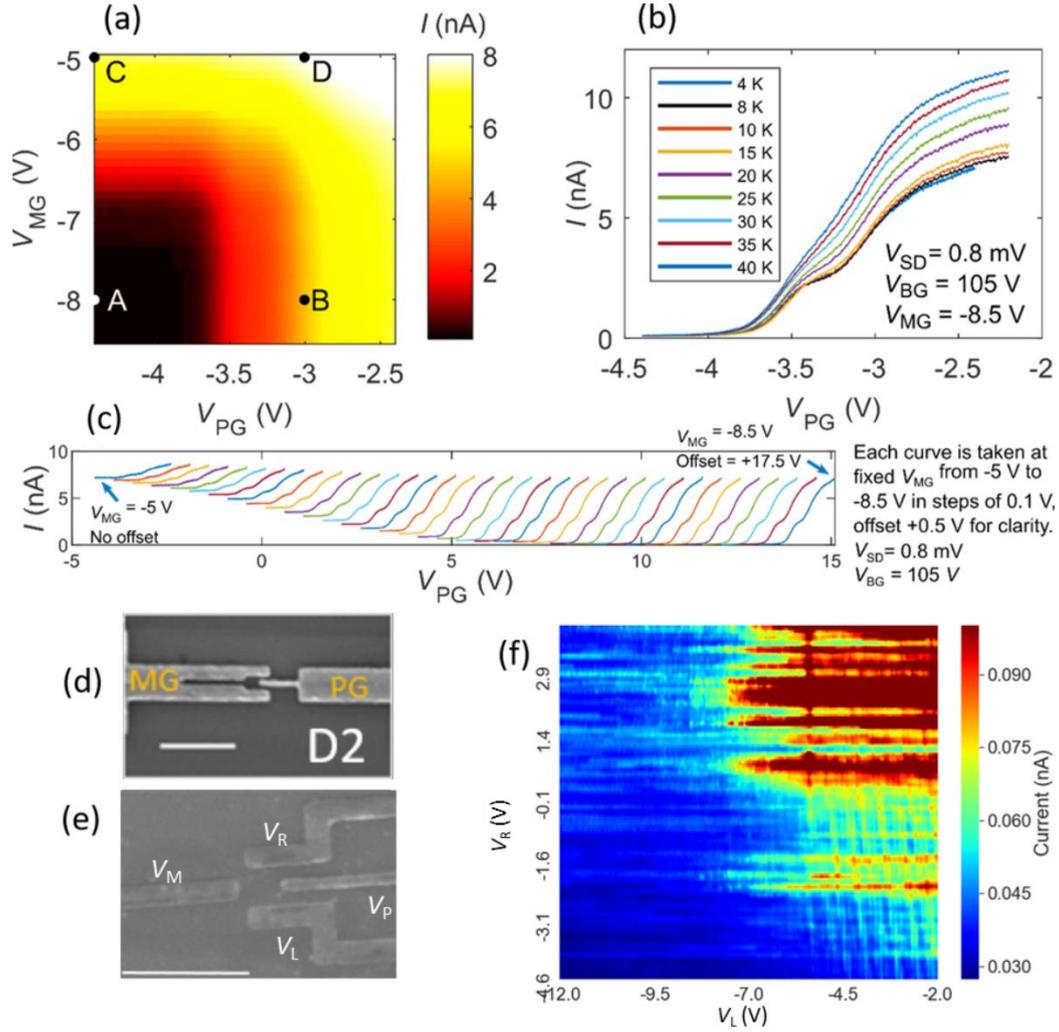

**Figure 9.** Electrical measurements of exfoliated few-layer WSe$_2$ and Chemical Vapor Deposition (CVD) grown single-layer MoS$_2$ devices. (a) Current through the WSe$_2$ device with split top gates on an Al$_2$O$_3$ dielectric, showing independent gate control over current pinch off. (b) A horizontal cut of (a) taken at $V_{MG}$ = -8.5 V showing a distinct current step that is increasingly smeared out at higher temperatures. (c) Multiple 1D cuts of (a) at different $V_{MG}$. SEM images of the (d) exfoliated few-layer WSe$_2$ device measured in (a-c) and the (e) CVD single-layer MoS$_2$ device with top confinement gates patterned on a HfO$_2$ dielectric. Scale bars are 500 nm. (f) Current through the device in (e) as a function of $V_R$ and $V_L$. $V_M$ and $V_P$ were set to -15 V. Multiple horizontal and vertical resonances are observed, indicating a disordered 2DEG. Reproduced with permission.[39] 2019, Springer Nature.

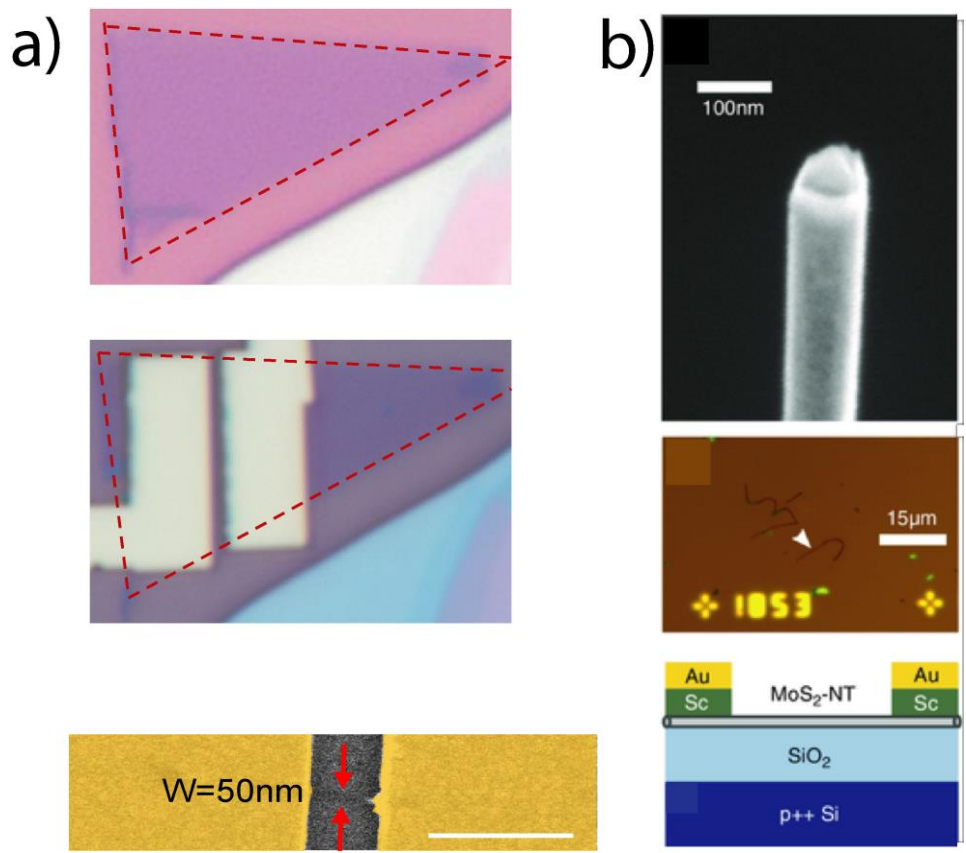

**Figure 10.** Fabrication process for a nanoribbon and nanotube based field effect transistors in single layer $MoS_2$. (a) Device fabrication of the nanoribbon transistor showing complete process from a $MoS_2$ flake (triangular purple region) to nanoribbon device using $SF_6$ dry plasma with metallic contacts shown as yellow stripes. The scale-bar in the bottom panel of (a) is 1 µm. Reproduced with permission.[108] 2019, American Chemical Society. (b) Nanotube based device grown by chemical transport reaction using iodine as a transport agent. Reproduced with permission.[105] 2019, WILEY-VCH.

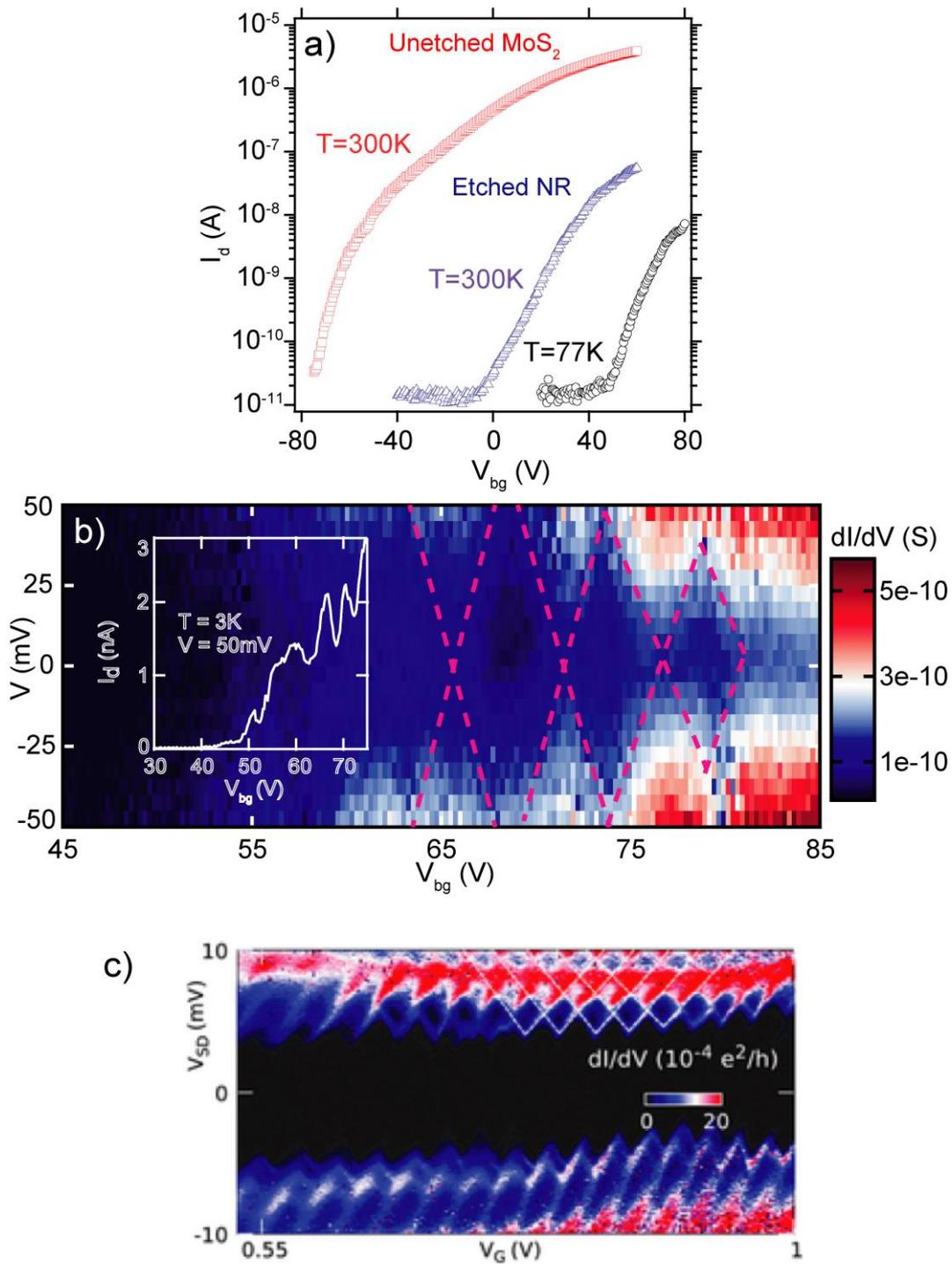

**Figure 11.** Electrical characterization of nanoribbon and nanotube based devices. (a) Drain current ($I_d$) vs $V_{bg}$ at V=100mV for unetched $MoS_2$ flake and nanoribbon at different temperatures. Graphs of $I_d$ vs $V_{bg}$: "□" for unetched $MoS_2$ flake at room temperature, "Δ" for etched nanoribbon at room temperature, and "O" for etched nanoribbon at $T$=77 K. (b) 2D conductance plot ($dI/dV$) as a function of $V$ (source-drain bias) and $V_{bg}$ (back-gate voltage) at $T$=3 K exhibiting Coulomb diamonds due to single electron transport through a quantum dot. Inset in (b) shows $I_d$ vs $V_{bg}$ at fixed $V$=5 mV and $T$=3 K. Reproduced with permission.[108] 2019, American Chemical Society. (c) Differential conductance ($dI/dV$) plot measured at 300mK as a function of source-drain bias and back gate voltage in nanotube device. Reproduced with permission.[105] 2019, WILEY-VCH.

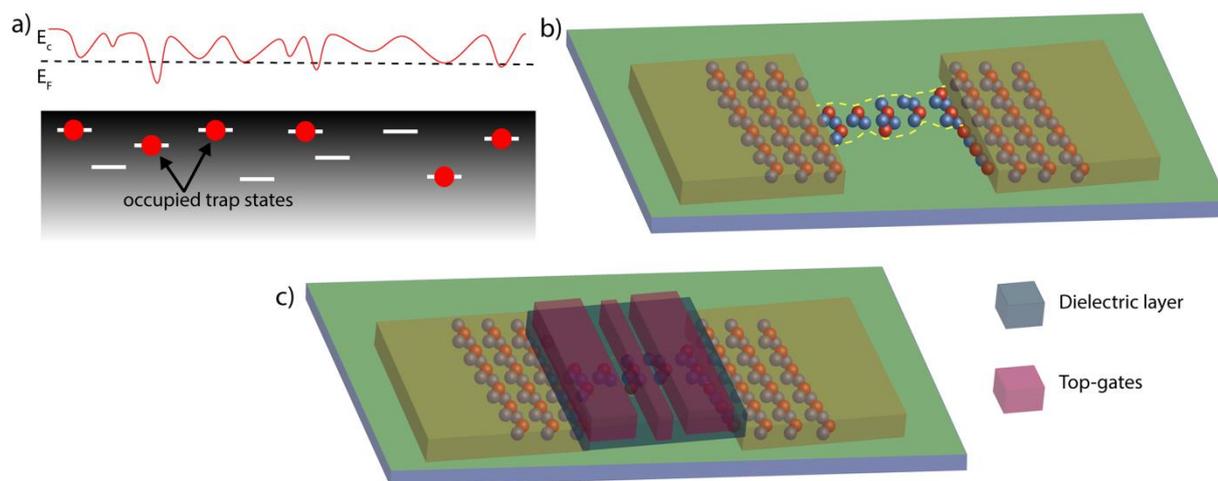

**Figure 12.** Origin of disorder in nanoribbon device and proposed architecture to overcome disorder. (a) Electrochemical potential fluctuations in MoS$_2$ due to trap states in SiO$_2$ substrate. (b) Artistic representation of non-uniformity in an etched nanoribbon device due to rough edge outlined by yellow dotted line resulting in formation of a quantum dot. (c) Schematic of locally gated nanoribbon device to tune out defect states and provide controlled quantum dot in nanoribbon (with local metallic top-gates separated from nanoribbon by a dielectric layer). Reproduced with permission.[108] 2019, American Chemical Society.